\documentclass[pdftex,twocolumn,epjc3]{svjour3}

\usepackage[utf8]{inputenc}

\usepackage{amsmath}
\usepackage{mathrsfs}
\usepackage{slashed}
\usepackage{dsfont}
\usepackage{graphicx}
\usepackage{subcaption}
\usepackage{float}
\usepackage{microtype}
\usepackage[colorlinks,citecolor=blue,urlcolor=blue,linkcolor=blue,bookmarks]{hyperref}
\usepackage[numbers,sort&compress]{natbib} 
\usepackage[margin=15pt,font=small,labelfont={bf,sf}]{caption}
\usepackage[dvipsnames,svgnames,table]{xcolor}

\newcommand{\muonetoJ}{\mu_1\mu_2\dots\mu_J}
\newcommand{\tr}{\operatorname{tr}}
\newcommand{\dbar}{\mathchar'26\mkern-11mu\mathrm{d}}

\journalname{Eur. Phys. J. A}

\begin{document}

\title{Spectra of light and heavy mesons with $J \le 5$  in a relativistic Bethe-Salpeter approach}

\author{Stephan Hagel\thanksref{addr1}, Christian S. Fischer\thanksref{e2,addr1,addr2}, Markus Q. Huber\thanksref{e3,addr1}, Jonathan Y. Yigzaw\thanksref{addr1}}

\thankstext{e2}{e-mail: christian.fischer@theo.physik.uni-giessen.de}
\thankstext{e3}{e-mail: markus.huber@theo.physik.uni-giessen.de}

\institute{Institut f\"ur Theoretische Physik, Justus-Liebig-Universit\"at Giessen, Heinrich-Buff-Ring 16, 35392 Giessen, Germany\label{addr1}
	\and
	Helmholtz Forschungsakademie Hessen f\"ur FAIR (HFHF), GSI Helmholtzzentrum f\"ur Schwerionenforschung, Campus Gie{\ss}en, 35392 Gie{\ss}en, Germany\label{addr2}
}

\date{\today}

\maketitle

\begin{abstract}
We extend the range of application of the relativistic Dyson-Schwinger/Bethe-Salpeter approach 
from previously discussed mesons with total angular momentum $J \le 3$ to the ones with $J=4,5$. 
On a technical level, the new element is the general Dirac tensor representations for the latter
which, to our knowledge, are presented here for the first time. As a first application, we provide 
an exploratory spectrum for these mesons in a rainbow-ladder truncation of Dyson-Schwinger and 
Bethe-Salpeter equations. We discuss the merits and limitations of this truncation and explore 
the shape of the heavy-quark potential corresponding to the underlying effective running coupling. 
With our predictions for the masses of ground state mesons with quantum numbers 
$J^{PC}=3^{--}, 4^{++}, 5^{--}$ we identify Regge trajectories in channels where the interaction model 
can be trusted on a semi-quantitative level. In other channels, discrepancies with experiments confirm the 
well-known need to go beyond rainbow-ladder in the Dyson-Schwinger/Bethe-Salpeter approach by using more 
sophisticated interactions.
\end{abstract}

\section{Introduction}\label{sec:Intro}

The study of the meson spectrum, in particular in the strange meson sector, is a high priority topic in
the COMPASS and the upcoming AMBER experiments \cite{COMPASS:2025wkw,Quintans:2022utc,Wallner:2022scd}. In these 
experiments, ground and excited states with various quantum numbers up to $J=5$ for kaons and even $J=6$ 
for light mesons can be extracted, thus offering the tantalizing possibility of a comprehensive picture of 
nature's light quark-antiquark states. 

The theoretical exploration of many of these states is hampered by the possibility of mixing between conventional 
quark-antiquark states, hybrid mesons that contain one or more constituent gluons, as well as more complex ones 
such as glueballs and four-quark states. This may be particularly true for the light meson sector, where a huge 
amount of literature is available dealing with this problem. Relativistic quark models, effective chiral 
Lagrangians, Hamiltonian approaches, QCD sum rules, Dyson-Schwinger and functional renormalization group methods 
as well as lattice QCD are methods of choice, see \emph{e.g.}~\cite{Brambilla:2014jmp} for a review and a guide
to further reading. In this work we concentrate on the functional approach via Dyson--Schwinger 
equations (DSEs) and Bethe--Salpeter equations (BSEs) which offers a direct connection between the details of the
non-perturbative quark-gluon interaction and the relativistic and field-theoretical description of bound-states.

In this framework, the Dirac structure of mesons is expressed in terms 
of complete sets of tensors representing all possible internal 
relativistic spin and orbital angular momentum combinations with fixed total
angular momentum $J$. These representations
have been worked out long ago for mesons with $J=0,1,2$ \cite{Joos:1962qq,Weinberg:1964cn,Zemach:1968zz,Krassnigg:2010mh}
and has been extended to $J=3$ in \cite{Fischer:2014xha}. Here, we further generalize these representations
to mesons with $J=4$ and $J=5$. This is particularly interesting since it allows us to extend previous studies
of Regge trajectories in the light and heavy meson sectors to higher spins in direct comparison to relativistic 
quark models that rely on linear rising (quasi-)potentials, see \emph{e.g.}~\cite{Godfrey:1985xj,Ebert:2009ub}.
 
For simplicity, and in order to connect to previous results in this framework, we start our investigation
using a simple rainbow-ladder framework with a flavour-diagonal vector-type interaction. Such a truncation
has the merit of preserving chiral symmetry thus reproducing important QCD constraints such as the 
(pseudo-) Goldstone boson nature of pseudoscalar mesons. It is however well-known \cite{Qin:2011xq,Blank:2011ha}
that the quality of this truncation type varies dramatically with quantum numbers. In this work we 
will confirm previous results on this topic for $J=0,1,2,3$ and add results for $J=4,5$. The necessary
extension of our study beyond rainbow-ladder, e.g. along the lines of \cite{Williams:2015cvx} is relegated to
future work. 

The paper is organized as follows. In Sec.~\ref{sec:DSEBSE} we introduce the framework of the 
DSEs and BSEs together with a discussion of the rainbow-ladder truncation and the model interactions 
employed. Aspects of the corresponding potentials are discussed in Sec.~\ref{sec:potentials}.
In Sec.~\ref{sec:results} we display our results. We discuss the resulting
spectra for light mesons, strange mesons ($S=0,1$) and heavy quarkonia 
and display corresponding Regge trajectories. We conclude in Sec.~\ref{sec:conclusion}.
Technical details, such as the extrapolation of our eigenvalue curves as well as the construction of the new
representations for $J=4,5$, are relegated to appendices.

\section{Dyson-Schwinger and Bethe-Salpeter equations}\label{sec:DSEBSE}

The basis of this work is the framework of functional methods, more specifically the Dyson-Schwinger
and Bethe-Salpeter equations, see e.g. \cite{Cloet:2013jya,Eichmann:2016yit,Eichmann:2025wgs,Huber:2025cbd}
for recent review articles. In this work, we need the solution of the quark propagator's DSE as input for 
the meson BSE to compute the composite state Bethe-Salpeter amplitudes.

\subsection{Quark propagator Dyson-Schwinger equation}\label{subsec:dse}

The DSE for the dressed quark propagator $S^{-1}(p) = i\slashed p A(p^2) + B(p^2)$ with dressing functions
$A(p^2)$ and $B(p^2)$ and $k=p-q$ reads\footnote{We use the notation $\dbar^n k= \mathrm{d}^n k / (2 \pi)^n$.}
\begin{multline}\label{eq:quark-dse} 
	S^{-1}(p) = Z_2 S_0^{-1}(p) + Z_{1F} g^{2}C_f \times \\
	\int \dbar^{4}q\,\Gamma_{qg}^{\mu}(k;-p,q)\,S(q)\,\gamma^{\nu}\,D^{\mu\nu}(k),
\end{multline}
where
\begin{align}
D^{\mu\nu}(k) = T_k^{\mu\nu} \frac{Z(k^2)}{k^2}= \left(\delta^{\mu\nu} - \frac{k^\mu k^\nu}{k^2}\right)\frac{Z(k^2)}{k^2}
\end{align}is the gluon propagator in Landau gauge
and 
\begin{equation}
\Gamma^\mu_{qg}(k;p,q) = \sum_{i=1}^{12}h_i(k;p,q)\lambda_i^\mu(k;p,q)
\end{equation}
denotes the quark-gluon vertex with Dirac tensors $\lambda_i^\mu$ and dressing functions $h_i$. 
Here and in the following we suppress all colour and flavour indices for simplicity. The bare quark propagator 
$S^{-1}_0(p)$ is obtained from the dressed one by replacing $A(p^2)=1$ and $B(p^2) = m_0$ with the bare quark mass 
$m_0$ related to the renormalized quark mass $m_q$ via $\,m_0=Z_m\, m_q$.
$Z_2$ and $Z_m$ are renormalization constants. The corresponding renormalization factor for the quark-gluon vertex is denoted by
$Z_{1F}$. The Casimir $C_F=4/3$ stems from the colour trace with $N_c=3$
and $g$ is the renormalized coupling constant of QCD.

\begin{table*}
	\begin{center}
		\begin{tabular}{c|cccc}
			\hline
			\hline
			& light & strange & charm   & bottom  \\
			\hline
			$m_q$     & $0.0037$ & $0.085$ & $0.830$ & $3.765$  \\
			$\eta$    & $1.8$     & $1.8$   & $1.157$  & $1.1$ \\
			$\Lambda$ & $0.72$    & $0.72$  & $0.72$  & $0.72$  \\
			\hline
			$\eta_\text{QC}$  &   $1.741$     &    &   &  \\
			$\Lambda_\text{QC}$ & $0.696$    &   &  &   \\
			\hline
			\hline
		\end{tabular}
	\end{center}
	\caption{
		Model and QCD parameters used for rainbow-ladder calculations.
		All quark masses and the parameter $\Lambda$ are given in units of GeV.
	}
	\label{tab:rl-parameters}
\end{table*}

The gluon propagator and the vertex obey their own set of DSEs which in turn introduce higher $n$-point functions, 
leading to an infinite tower of coupled integral equations.
In general, it cannot be solved in closed form.
This makes it necessary to truncate the equations at some level.
Elaborate and self-contained schemes have been developed based on truncating the system to the primitively divergent correlation functions.
For pure gauge theory \cite{Huber:2020keu,Huber:2018ned}, the spectrum of glueballs was calculated up to $J=4$ \cite{Huber:2020ngt,Huber:2021yfy}.
The results support what is called apparent convergence, i.e., they do not change significantly when including higher $n$-point functions \cite{Huber:2025kwy}.
A not yet fully self-contained version of such a scheme has also been applied to the light meson spectrum \cite{Williams:2015cvx}.

In this exploratory work, however, we resort to a much simpler scheme. The so-called rainbow-ladder truncation has 
been widely used in the past. It is technically easy to handle and is known to preserve both the vector and axialvector
Ward-Takahashi identities and thus the effects of spontaneous chiral symmetry breaking.
In this truncation, the full quark-gluon vertex is reduced to its leading (tree-level) tensor structure
$\Gamma^\mu_{qg}(k;p,q)=Z_{1F}h_1(k^2)\gamma^\mu$, and the momentum
dependence of the dressing function $h_1(k^2)$ is reduced to the gluon
momentum $k^2$. This dressing function can then be combined with $Z(k^2)$
from the gluon propagator and a subset of the renormalization factors 
into an effective, renormalization group invariant running coupling $\alpha_\text{eff}(k^2)$. The truncated quark DSE then reads
\begin{multline}\label{eq:quark-dse-truncated}
	S^{-1}(p) = Z_2 S_0^{-1}(p) + C_{f}Z_2^{2}4\pi \times \\
	\int \dbar^{4}q\gamma^{\mu}S(q)\gamma^{\nu}T_k^{\mu\nu}\frac{\alpha_\text{eff}(k^2)}{k^2}.
\end{multline}
By projecting onto the tensor structures of the quark propagator, one obtains two coupled scalar integral equations for the
dressing functions $A(p^2)$ and $B(p^2)$ which can be solved numerically.

There are several potential choices for the effective running coupling $\alpha_\text{eff}(k^2)$.
Two closely related ones with the most widespread use have been proposed by Maris and Tandy (MT)~\cite{Maris:1999nt}
and by Qin et al. (QC)~\cite{Qin:2011dd}. Their general form is given by 
\begin{equation}\label{eq:effective-coupling}
\alpha_\text{eff}(k^2) = \alpha_\text{IR}(k^2) + 
\frac{\pi\gamma_m\left( 1 - \mathrm{e}^{-k^2/\Lambda_0^2} \right)}
{\ln\sqrt{\mathrm{e}^2-1+\left( 1+k^2/\Lambda_\text{QCD}^2 \right)^2}}
\end{equation}
with the large-momentum running dictated by perturbation theory with $\gamma_m=12/(11N_c -2 N_f)$, $N_c=3$ and $N_f=4$. 
The corresponding scales are given by $\Lambda_0=1\,$GeV and $\Lambda_\text{QCD}=0.234\,$GeV \cite{Maris:1999nt}.
At small momenta, the form of the couplings mainly differ by a factor of $k^2/\Lambda^2$:
\begin{align}
	\alpha_\text{IR,MT}(k^2) &= \pi \eta^7 \,\frac{k^4}{\Lambda^4}\, \mathrm{e}^{-\eta^2 k^2/\Lambda^2}
	\label{eq:maris-tandy}\\
	\alpha_\text{IR,QC}(k^2) &= \pi \eta_\text{QC}^7 \,\frac{k^2}{\Lambda_\text{QC}^2}\, \mathrm{e}^{-\eta_\text{QC}^2 k^2/\Lambda_\text{QC}^2}
	\label{eq:qin-chang}
\end{align}
Here, we use a parametrization of the QC model that is as close to the MT model as possible.
For reference, we give the connection to the parameters of \cite{Qin:2011dd}: $D_\text{QC}= \eta_\text{QC}^3\Lambda_\text{QC}^2/2$ and $\omega_\text{QC} = \Lambda_\text{QC}/\eta_\text{QC}$.
We will compare both models to some extent throughout this work.

The values for the model parameters of the interactions together with the values for the renormalized 
quark masses $m_q$ are detailed in Tab.~\ref{tab:rl-parameters}. The scale $\Lambda$ is determined
via matching with the experimental value for the pion decay constant. The quark masses are given at the scale
$\mu = 19$ GeV in a $\widetilde{\text{MOM}}$-subtraction scheme. The light quark mass is fixed such that the 
experimental charged pion mass is reproduced, the strange quark mass is adapted to the charged kaons, the
charm quark mass to the $J/\Psi$ and the bottom mass to the $\Upsilon(1S)$. 

At the end of this subsection, let us briefly discuss what to expect from a pure rainbow-ladder interaction. 
The dressed quark-gluon vertex contains twelve different tensor structures in total, and features interaction 
strength not only in its $\gamma_\mu$-part, but also in other tensor structures. Depending on the choice of 
basis, there are at least two additional components that are as important as $\gamma_\mu$, see e.g.  
\cite{Mitter:2014wpa,Williams:2015cvx,Aguilar:2024ciu,Gao:2021wun} for details. In a rainbow-ladder type interaction, these
are neglected and interaction strength is shifted to the $\gamma_\mu$-part. As a result, such an 'approximation' 
provides still enough interaction to trigger dynamical chiral symmetry breaking, but details on the spin structure
of the interaction are lost. This has a drastic effect on the resulting meson spectra as we will discuss below. 
For calculations of meson spectra beyond rainbow ladder see e.g. \cite{Chang:2011ei,Williams:2015cvx,Ferreira:2026gbe}. 

\subsection{Bethe-Salpeter equation}\label{subsec:bether-salpeter-equation}

In order to investigate mesons as bound states and resonances of quarks and antiquarks, we use the Bethe-Salpeter equation
\begin{multline}
	[\Gamma^{\muonetoJ}(p, P)]_{tu} = \int\dbar^4 q K_{tu,rs}(q, p, P) \\
	[S(q_+) \Gamma^{\muonetoJ}(q,P) S(q_-)]_{sr} .
\end{multline}
The quantity $\Gamma^{\muonetoJ}(q,P)$ is the Bethe-Salpeter amplitude and $K(q,p,P)$ is the quark-antiquark scattering
kernel with generic indices for colour, flavour and Dirac structures.
Different quantum numbers of bound states can be investigated by constructing the tensor basis for the Bethe-Salpeter
amplitude $\Gamma^{\muonetoJ}(q,P)$ such that it has the corresponding transformation properties.
\ref{sec:app:bases} contains details on how a basis can be constructed accordingly.

The BSE can be solved numerically as eigenvalue equation of the form
\begin{equation}
	(KG)\cdot\Gamma = \lambda(P^2)\, \Gamma \,,
\end{equation}
where at the physical meson mass the eigenvalue is equal to one, i.e. $\lambda(P^2 = -m^2) = 1$.
Depending on the mass of the meson, the kernel can be evaluated at the required value of $P^2$.
For heavier mesons, however, this is not directly possible due to singularities in the quark propagator, see \ref{sec:app:extrapol}.
In that case, we perform the calculation for safe values of $P^2>-m^2$ and extrapolate the eigenvalue curve to the physical point as also discussed in \ref{sec:app:extrapol}.
Via a sampling of the points used for the extrapolation we determine an error estimate which is shown as error bar in the plots.

The scattering kernel in the Bethe-Salpeter equation cannot be chosen arbitrarily, but it has to match the truncation
scheme used in the DSE such that the axialvector Ward-Takahashi identity is satisfied.
As a consequence, the (pseudo-)Goldstone nature of the pion is preserved. Corresponding details and an explicit derivation of this property can be found, e.g. in Ref.~\cite{Eichmann:2016yit,Eichmann:2025wgs}.
For the rainbow-ladder truncation, this is achieved by using a single effective gluon exchange between two bare vertices
\begin{equation}\label{eq:rl-kernel}
	K_{ab,cd}(p,q,P) = -C_f Z_2^2 4\pi [\gamma^{\mu}]_{ab} [\gamma^\nu]_{cd} T_{\mu\nu}(k)
	\frac{\alpha_\text{eff}(k^2)}{k^2}.
\end{equation}
It is of utmost importance that the effective running coupling used is the same as the one in the DSE for the quark
propagator, as even slight deviations cause the pion to generate a sizeable mass in the chiral limit.

\begin{figure*}
	\begin{subfigure}{0.495\linewidth}
		\includegraphics[width=\linewidth]{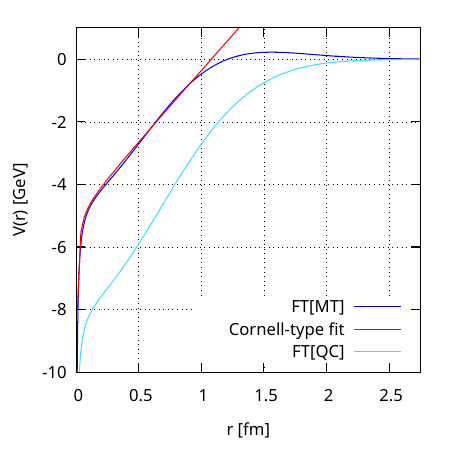}
	\end{subfigure}
	\begin{subfigure}{0.495\linewidth}
		\includegraphics[width=\linewidth]{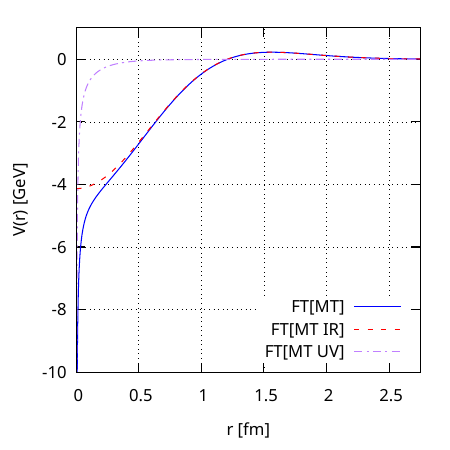}
	\end{subfigure}
	\caption{\textbf{Left panel:} 
		Interaction potentials obtained via Fourier transformation of effective running couplings.
		The MT potential has a (positive) maximum around $r=1.6$\,fm and approaches zero from the positive side
		while the QC potential approaches zero monotonously.
		\textbf{Right panel:} The contributions of the infrared and ultraviolet parts of the effective running coupling to
		the MT potential.
	}
	\label{fig:potentials1}
\end{figure*}

\section{Interaction potentials}\label{sec:potentials}

In the heavy quark sector, i.e. for bottomonia and to a lesser degree charmonia, the notion of an interaction potential
between the quark-antiquark pair of a conventional meson is adequate, see e.g. \cite{Bali:2000gf} for a review. 
Such potentials are routinely used in NRQCD \cite{Brambilla:1999xf,Brambilla:2000gk} and have been investigated 
also in lattice QCD \cite{Koma:2006si,Koma:2006fw,Eichberg:2024svw}. In its simplest form, i.e. without spin and 
angular momentum dependent parts, such a potential can be parametrized by the Cornell ansatz
\begin{equation}\label{eq:cornell}
	V(r) = -\frac43 \frac{\alpha_s}{r} + r\, \sigma(r)
\end{equation}
with the strong coupling constant parameter $\alpha_s$ and a constant string tension $\sigma(r) \approx \sigma_0$ 
for small values of $r$, while $\sigma(r) \rightarrow 1/r$.
In the limit  $r\rightarrow \infty$, $\sigma(r) \rightarrow 1/r$ implements a smoothed out version of the effects of string breaking~\cite{Bali:2005fu}.

A striking feature of the QCD spectrum is the appearance of approximately linear Regge trajectories in Chew-Frautschi 
plots of squared mass vs. total angular momentum. While such a behaviour is very easily explained in simple models like 
a spinning stick or string picture, it has to emerge from the underlying complex dynamics of QCD \cite{Greensite:2011zz}. 
Of course, the rainbow-ladder interaction model discussed above will hardly contribute fundamental insights 
into this issue. Nevertheless, as we will see, it is interesting to compare its corresponding heavy-quark 
interaction potential and the resulting meson spectra with these general expectations.    

In order to obtain a potential from such a running coupling, one has to perform a Fourier transformation with respect to the three-dimensional momentum of the truncated scattering kernel used in the BSE, see, e.g.,~\cite{Cucchieri:2017icl}.
Performing such a transformation on a bare (massive) single boson exchange kernel, one obtains a Yukawa potential
which in the limit $m\rightarrow 0$ becomes a Coulomb potential. Applying the same procedure to the rainbow-ladder
kernel~(\ref{eq:rl-kernel}) with any effective running coupling which only depends on the squared exchange 
momentum corresponds to evaluating the integral 
\begin{align}
	V(\vec r) &= - C_f 4\pi \int \dbar^3 k \mathrm{e}^{-i \vec k \cdot \vec r} \frac{\alpha(\vec{k}^2)}{\vec{k}^2}.
\end{align}
With spherical coordinates, two of the integrals can be done analytically, the remaining one is performed
numerically.

The resulting interaction potentials for the effective running couplings in Eq.~(\ref{eq:effective-coupling}) are displayed in the left diagram of Fig.~\ref{fig:potentials1} for the models (\ref{eq:maris-tandy}) and (\ref{eq:qin-chang}). It is 
interesting to note that the qualitative behaviour of both potentials follows the generic QCD 
form of a Coulomb term plus a linear rising potential and includes a transition region to a 
constant long distance behaviour. Furthermore, in the right diagram of Fig.~\ref{fig:potentials1} we
show that the ultraviolet logarithmic part of the effective running coupling is responsible for the
short distance Coulomb type behaviour of the potentials, whereas the infrared exponential parts of the
effective running couplings generate the non-trivial parts at intermediate distances. 

In order to discuss the physics of these potentials, it is instructive to compare our results to the 
ones of Ref.~\cite{Cucchieri:2017icl}, where a corresponding potential from a quenched lattice gluon
propagator has been extracted (see their Fig.2). The short distance behaviour in both approaches is 
generated by the perturbative one-gluon exchange modified with logarithmic corrections of the gluon
(Ref.~\cite{Cucchieri:2017icl}) or the gluon combined with logarithmic corrections of the quark-gluon
vertex (our potential). Both types of corrections do not modify the Coulomb part
substantially, as expected. The large distance behaviour of both approaches, in particular when 
compared to our potential from the MT interaction, is also very similar. The underlying reason in both
cases is that the dressing function of the gluon (Ref.~\cite{Cucchieri:2017icl}) as well as the effective
running coupling (in our case) vanish at zero momentum. Consequently, the long range potential in both
cases goes to zero. It has been noted in Ref.~\cite{Cucchieri:2017icl} that the physics of a long distance
linear behaviour of the potential associated with confinement in pure Yang-Mills theory and the phenomenon 
of string breaking in QCD is not included in the simple one-gluon exchange picture. This is also true for
our approach: the simple models for the effective running coupling used in this work do not include this 
type of physics.\footnote{Note, however, that in principle the combination of non-perturbative dressings of 
gluon and quark-gluon vertices in a fully non-perturbative calculation of the quark-antiquark potential 
may show these effects, as argued in Refs.~ \cite{Alkofer:2006gz,Alkofer:2008tt}: the dressings may 
conspire to generate an overall $1/k^4$-term in the potential which suffices to create an asymptotic 
linear rising part \cite{West:1982bt}.}
\begin{figure*}
	\centering
	\includegraphics[width=0.7\linewidth]{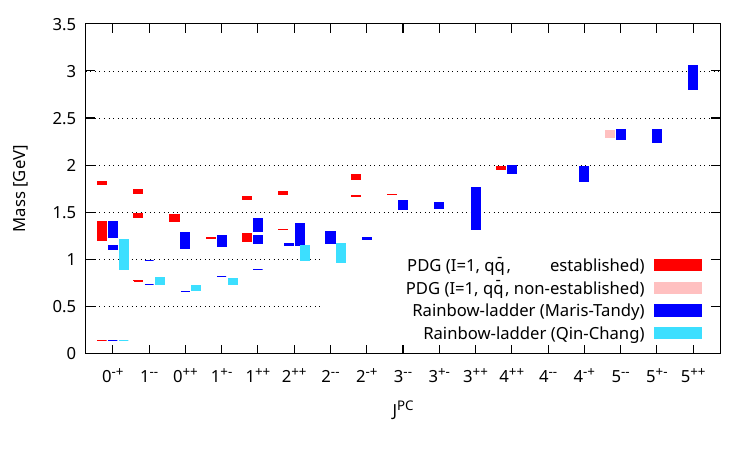}
	\caption{Spectrum of $q\bar q$ states obtained from a rainbow-ladder truncation with the Maris-Tandy 
		and Qin-Chang models compared to the experimental spectrum in the isovector channel~\cite{ParticleDataGroup:2024cfk}.}
	\label{fig:spectrum-mt-light}
\end{figure*}

A significant difference between our potential and the one of Ref.~\cite{Cucchieri:2017icl} is the appearance
of an approximately linear rising part at intermediate distances for the potential from the MT model. 
Adding a constant term to the expression of the Cornell-potential, Eq.(\ref{eq:cornell}), we obtain a good
fit of our numerical data for $\alpha_S = 0.162$ for the strong coupling parameter, a constant string tension 
of $\sigma = 0.890\,\text{GeV}^2$ and a constant offset of $-4.84$ GeV. This is shown as 'Cornell-type fit'
in Fig.~\ref{fig:potentials1}. The appearance of an almost linear region at intermediate distances has interesting
consequences since it allows for the appearance of (approximate) Regge trajectories in the calculated spectrum.
We will come back to this point in the next section. Also note that the potential obtained from 
the QC model shows a slight bend in the linear region. Changing the linear term in our fit to be of 
the form $\sigma_0 r^n$ and using $n$ as an additional fitting parameter the fit prefers a clear deviation 
from linear behaviour with $n=1.33$. It is thus not clear whether the QC model is similarly supportive 
of Regge behaviour as the MT model. This will also be discussed below.  

We wish to emphasize that the potentials calculated in this section are for illustrational purposes only
and we do not use them to determine the results shown below. Instead, as described in section \ref{sec:DSEBSE},
we solve the fully relativistic Dyson-Schwinger and Bethe-Salpeter equations in momentum space and without 
recurse to any non-relativistic approximations.  

\begin{figure*}
	\centering
	\includegraphics[width=0.7\linewidth]{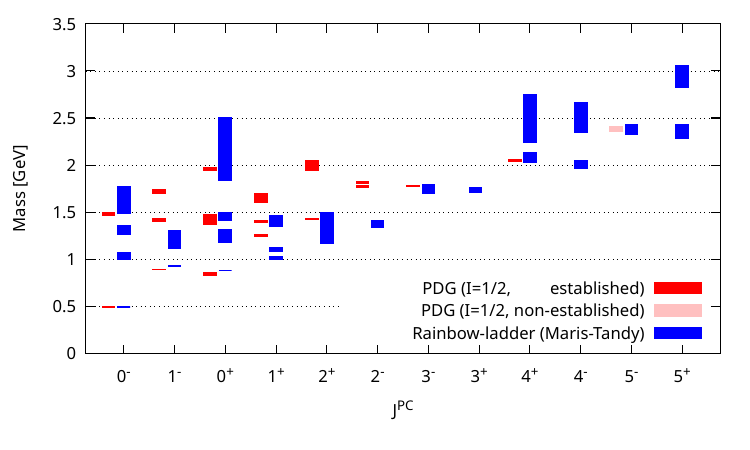}
	\caption{Spectrum of states with one light and one strange (anti-)quark obtained from a rainbow-ladder truncation
		with the Maris-Tandy model compared to the experimental kaon spectrum~\cite{ParticleDataGroup:2024cfk}.}
	\label{fig:spectrum-mt-kaon}
	\centering
	\includegraphics[width=0.7\linewidth]{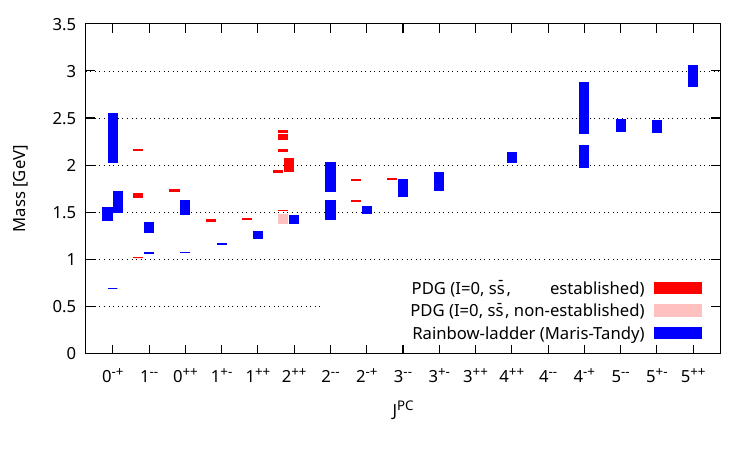}
	\caption{Spectrum of $s\bar s$ states obtained from a rainbow-ladder truncation with the Maris-Tandy model 
		compared to the	experimental isoscalar spectrum~\cite{ParticleDataGroup:2024cfk}.}
	\label{fig:spectrum-mt-ssbar}
\end{figure*}
\begin{figure*}
	\centering
	\includegraphics[width=0.7\linewidth]{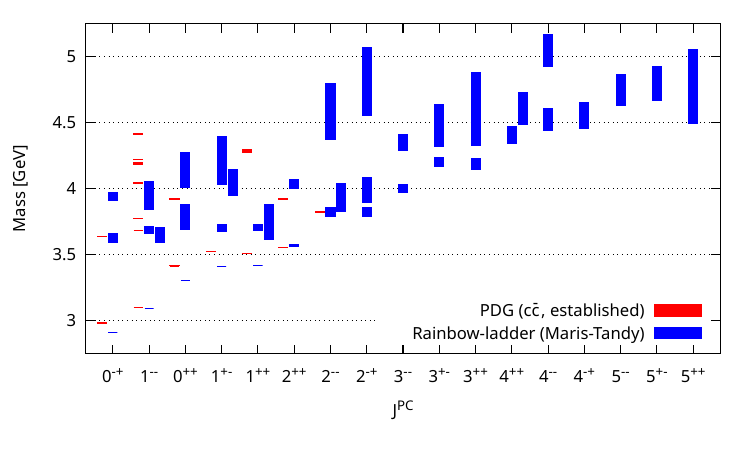}
	\caption{
		Charmonium spectrum obtained from a rainbow-ladder truncation with the Maris-Tandy model 
		and $\eta = 1.16$ compared to experimental data~\cite{ParticleDataGroup:2024cfk}.
		Note that potential tetraquark candidates such as the $\eta_{c1}(3872)$ are not included.}
	\label{fig:spectrum-mt-charmonium}
	\centering
	\includegraphics[width=0.7\linewidth]{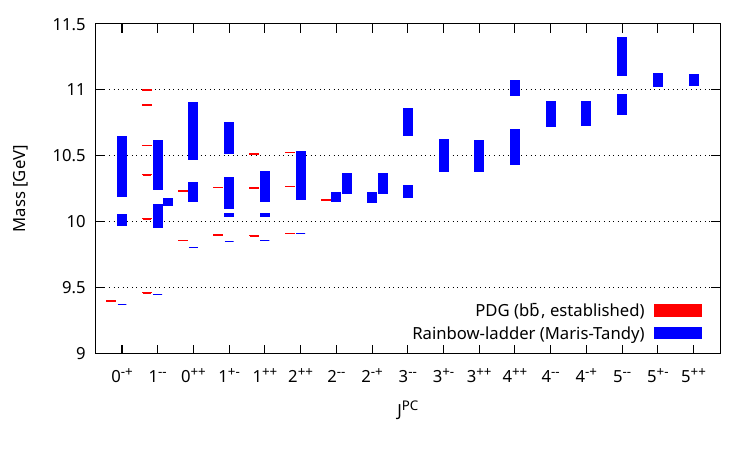}
	\caption{
		Bottomonium spectrum obtained from a rainbow-ladder truncation with the Maris-Tandy model
		and $\eta = 1.1$ compared to experimental data~\cite{ParticleDataGroup:2024cfk}.
	}
	\label{fig:spectrum-mt-bottomonium}
\end{figure*}

\section{Results}\label{sec:results}
\subsection{Meson spectra}\label{subsec:results:meson}

The results for the spectrum of mesons composed of light quarks is shown in Fig.~\ref{fig:spectrum-mt-light}. 
For a detailed discussion of the light meson rainbow-ladder spectrum we refer the reader to Ref.~\cite{Fischer:2014xha}. 
Here we focus on general aspects and in particular sequences including the new states with larger total angular momentum quantum numbers. 

As described at the end of section \ref{subsec:dse}, the pion mass has been used as input to determine the
light quark masses. The first non-trivial result is then the hyperfine splitting of pseudo-scalar and vector 
meson ground states, which comes out about right \cite{Maris:1999nt}.\footnote{Note that a general feature of rainbow-ladder 
type truncations is the absence of thresholds and dynamically generated decay widths in the BSEs. Nevertheless, 
for states with a decay width (imaginary part of the pole position) much smaller than the real part of the pole position, the real part can be extracted and decay constants can be calculated reliably via triangle diagrams, see, e.g., \cite{Jarecke:2002xd} for corresponding results for the $\rho$-meson.
More advanced kernels including $\pi \pi$ intermediate states have been employed in 
Refs.~\cite{Williams:2018adr,Santowsky2020,Miramontes:2021xgn} and have been shown to produce the two-pion right-hand cut in the
iso-vector and iso-scalar channels.}
The light scalar and axialvector channels are known to be not well represented in rainbow-ladder 
type kernels for different reasons. In the scalar channel, the ground state is not a conventional quark-antiquark state, but a four-quark
state with a strong coupling to isoscalar $\pi \pi$ intermediate channels and only very small quark-antiquark contributions. 
This has been demonstrated by many approaches (see e.g. \cite{Pelaez:2015qba} for a review) but also in the DSE/BSE framework 
in a series of works \cite{Heupel:2012ua,Eichmann:2015cra,Santowsky2020}.
In the axial-vector channel the rainbow-ladder masses for the ground state are too small, mainly due to the omission of
the other tensor structures of the quark-gluon vertex. Once these are included, the masses of the light axialvectors
come out about right \cite{Heupel:2014ina}. For the other mesons with $J>1$, no beyond rainbow-ladder calculations
are available so far. Comparing our results using the Maris-Tandy interaction with the available experimental data 
in these channels, it appears as if the sequence of ground states lying on the Regge trajectory with quantum numbers 
$J^{PC}=1^{--}, 2^{++}, 3^{--}, 4^{++}, 5^{--}$ is in good agreement with experiment (with slight tensions in the 
$2^{++}$-channel). This result has already been anticipated in Ref.~\cite{Fischer:2014xha} on the basis of states with 
$J \le 3$. Our new result for the $J^{PC}=4^{++}$ confirms this notion and we predict a $J^{PC}=5^{--}$ state in the mass
region around $2325\pm52.$ MeV.

In contrast, states on the Regge trajectory with quantum numbers $J^{PC}=1^{+\pm}, 2^{-\pm}, 3^{+\pm}, 4^{-\pm}, 5^{+\pm}$ 
are not well described within the rainbow-ladder truncation, indicating that further tensor structures in the interaction 
kernel are mandatory to describe those states. A further general aspect clearly visible in the spectrum is the poor 
description of radially excited states with the rainbow-ladder truncation. Also here, going beyond the rainbow improves the situation
drastically \cite{Heupel:2014ina}.

Finally, comparing results from the Maris-Tandy model with those obtained with the 
Qin-Chang interaction,
we find that both versions of the effective running coupling deliver qualitatively similar results with only slight advantages
of MT in the $2^{++}$ tensor channel. Overall, the ground states in QC for larger quantum numbers are somewhat smaller in mass
than the corresponding ones from the MT interaction. This effect may be understood by comparing the Fourier transforms
of the effective running couplings, i.e. the interaction potentials depicted in Fig.~\ref{fig:potentials1}. Higher mass 
states feel the (approximately) linearly rising parts of the potential.
The QC potential is much broader which results
in states that have a higher binding energy and thus lower masses compared to the MT potential. In our computations, we find 
that the MT coupling is numerically easier to handle. In fact, numerical problems with QC did prevent us from extracting a 
spectrum beyond $J=2$: we were not able to go as far into the complex momentum plane as with MT leaving us with not enough 
leverage for extrapolating to the physical point. Thus at present we cannot say whether the QC-interaction also supports
Regge-trajectories. Similar problems are encountered for heavier quarks and we therefore restrict ourselves to the MT 
interaction in the following.

A qualitatively similar result to the one for the light mesons can be seen for the kaon spectrum in
Fig.~\ref{fig:spectrum-mt-kaon}. Again, the ground states of the Regge trajectory 
$J^P=0^+, 1^-, 2^+, 3^-, 4^+, 5^-$~\footnote{Since $s\bar q$ states are not charge-conjugation eigenstates, 
the $C$-quantum number is not well-defined.} and the pseudoscalar ground state are in agreement with experiment (where
experimental data are available), while other quantum numbers and excited states are not. Our new result $2082\pm57$ MeV
for the $J^{PC}=4^{++}$ state agrees with the $K^*_4(2045)$ and we predict a $J^{PC}=5^{--}$ state in the mass
region around $2378$ MeV.
Notably, the kaonic state with quantum numbers $J^P = 0^+$ is described much better than the light quark state. 
This agreement may very well be accidental since we expect sizeable upwards shift of the masses of the
quark-antiquark states in a beyond rainbow-ladder truncation. A corresponding four-quark state has already been 
identified as the ground state in this channel \cite{Heupel:2012ua,Eichmann:2015cra,Santowsky2020} with a mass 
corresponding to the experimental $K_0^*(700)$ shown in the plot. It is also possible that in the $S=1$-sector the 
mixing between two- and four-quark states is much stronger than in the light quark sector. This needs to be explored
in future work. 

\begin{figure*}
	\centering
	\begin{subfigure}{0.45\linewidth}
		\includegraphics[width=\linewidth]{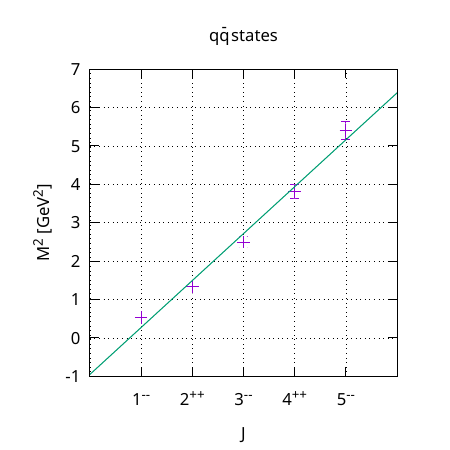}
	\end{subfigure}
	\begin{subfigure}{0.45\linewidth}
		\includegraphics[width=\linewidth]{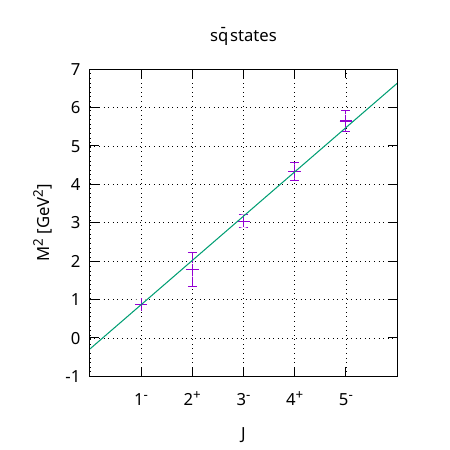}
	\end{subfigure}
	\\
	\begin{subfigure}{0.45\linewidth}
		\includegraphics[width=\linewidth]{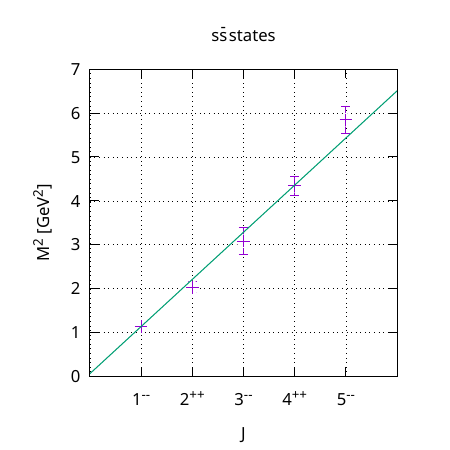}
	\end{subfigure}
	\begin{subfigure}{0.45\linewidth}
		\includegraphics[width=\linewidth]{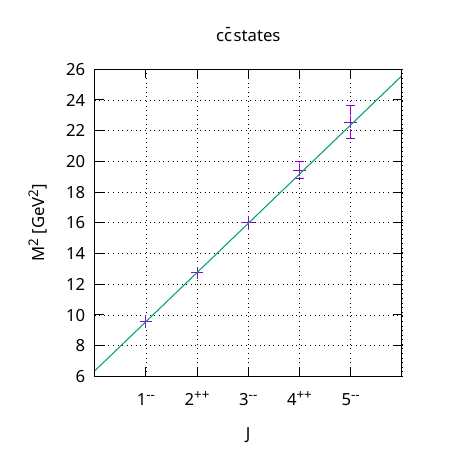}
	\end{subfigure}
	\\
	\begin{subfigure}{0.45\linewidth}
		\includegraphics[width=\linewidth]{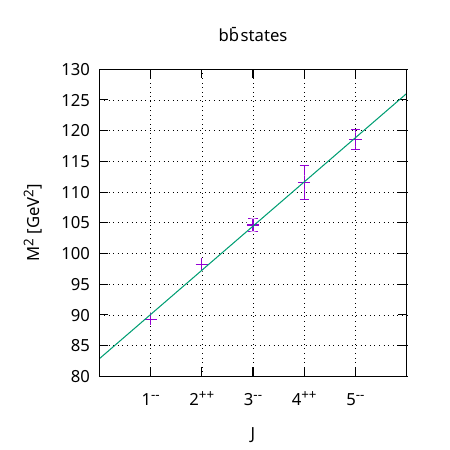}
	\end{subfigure}
	\caption{Chew-Frautschi plots for different quark contents.}
	\label{fig:chew-frautschi}
\end{figure*}

The $s\bar s$ spectrum is shown in Fig.~\ref{fig:spectrum-mt-ssbar}.
It looks qualitatively similar to the light quark results.
Here we do not compare to the experimental result for the $\eta'$ due to strong mixing in the pseudoscalar
channel and the topological mass effect of the $U_A(1)$ anomaly which is not captured in the rainbow-ladder
truncation. The interested reader may consult \cite{Alkofer:2008et} for a way to include the anomaly in the DSE/BSE
framework. In the other channels, mixing is much smaller and experimental states which are dominated by
their $s\bar s$ component, such as the $\phi$ states, may be cleanly identified. Again, in the sequence 
$J^{PC}=1^{--}, 2^{++}, 3^{--}, 4^{++}, 5^{--}$ the $\phi(1020)$ and $f_2(1430)$ (experimentally seen but 
not confirmed) and the $\phi_3(1850)$ are well reproduced by our calculation. We predict a state with $J^{PC}=4^{++}$
at $2084 \pm 53$ MeV and a state with $J^{PC}=5^{--}$ in the mass region around $m_{\phi_5}=(2418 \pm 64)$\,MeV.
\begin{table*}
	\begin{center}
		\begin{tabular}{c|ccccc}
			\hline
			\hline
			& $q\bar{q}$       & $s\bar{q}$       & $s\bar{s}$       & $c\bar{c}$      & $b\bar{b}$ \\
			\hline
			$M_0^2$   & $-0.95 \pm 0.28$ & $-0.29 \pm 0.03$ & $ 0.06 \pm 0.06$ & $6.35 \pm 0.02$ & $82.874 \pm 0.74$ \\
			$\beta$   & $ 1.22 \pm 0.09$ & $ 1.15 \pm 0.03$ & $ 1.07 \pm 0.06$ & $3.20 \pm 0.02$ & $ 7.193 \pm 0.22$ \\
			\hline
			\hline
		\end{tabular}
	\end{center}
	\caption{Parameters of Regge trajectories for $q\bar{q}$, $s\bar{q}$, $s\bar{s}$, $c\bar{c}$ and $b\bar{b}$ mesons with natural parity.}\label{tab:regge-fits}
\end{table*}

Fig.~\ref{fig:spectrum-mt-charmonium} shows our results for the charmonium spectrum with the model parameter $\eta$ fixed 
in Ref.~\cite{Fischer:2014cfa}.  
For heavy quarks, i.e. charm and bottom, we need to evaluate the corresponding quark propagators deep in the time-like 
region which introduces additional numerical uncertainties. In turn, the momentum region where we can evaluate the
Bethe-Salpeter equation safely does not exceed the lightest ground states and in general we have to extrapolate the 
eigenvalue of the Bethe-Salpeter matrix further towards larger masses. This leads to higher statistical uncertainties
and larger error bars, see \ref{sec:app:extrapol} for technical details. Nevertheless, we can again identify the same
pattern as for the light and strange quark sector. In the Regge sequence $J^{PC}=1^{--}, 2^{++}, 3^{--}, 4^{++}, 5^{--}$ 
the mass of the $J/\Psi$ is used as input to fix our charm quark mass and the experimental mass of the $\chi_{c2}(1P)$ is 
well reproduced. We predict states with $J^{PC}=3^{--}, 4^{++}, 5^{--}$ at $4001\pm33$~MeV, $4407\pm64$~MeV and $4749\pm115$~MeV,
respectively. Note that the PDG lists a $\psi_3(3842)$ state as potential $3^{--}$ candidate, with as yet unmeasured $J^P$, but
reference to quark model predictions. Since our predicted mass for this state is much higher, this identification might 
be called into question.  

The same situation can be seen in the bottomonium spectrum, Fig.~\ref{fig:spectrum-mt-bottomonium}. In the Regge sequence 
$J^{PC}=1^{--}, 2^{++}, 3^{--}, 4^{++}, 5^{--}$ the mass of the $\Upsilon(1S)$ is used as input to fix the bottom quark mass 
and the experimental mass of the $\chi_{b2}(1P)$ is well reproduced with our choice of model parameter $\eta$. Based on this
identification, we predict states with $J^{PC}=3^{--}, 4^{++}, 5^{--}$ at $10229\pm48$~MeV, $10563\pm133$~MeV and
$10889\pm75$~MeV, respectively. 

In general, when comparing results for the light, hidden strange, hidden charm and hidden bottom meson spectra, see
Figs.~\ref{fig:spectrum-mt-light},\ref{fig:spectrum-mt-ssbar},\ref{fig:spectrum-mt-charmonium},\ref{fig:spectrum-mt-bottomonium},
it is notable that the rainbow-ladder results for channels not on this trajectory become better with heavier quark masses: 
As noted above, there are substantial discrepancies between our results and the experimental masses in the scalar and 
axialvector channels in the light and strange meson sector. In the charmonium energy region, this situation improves 
drastically and in the bottomonium region it improves further. The underlying reason is the suppression of dynamical 
effects in the quark-gluon vertex with quark mass, which becomes exact in the limit of infinite quark mass and entails 
dominance of the leading tensor structure $\Gamma_\mu \sim \gamma_\mu$ of the vertex.  Since this is the (only) structure 
that is included in rainbow-ladder, truncations of this type improve in the heavy quark limit. In Coulomb gauge, this 
can even be shown analytically \cite{Watson:2012ht}.

\subsection{Regge trajectories}\label{subsec:results:regge}

Compared to experiment, we obtain reasonable results in channels on the natural Regge trajectories with quantum numbers
$J^{PC}=1^{--}, 2^{++}, 3^{--}, 4^{++}, 5^{--}$. 
Here, we collect these results and make a quantitative check for Regge behaviour.
This is of particular interest in the light of the discussion in Sec.~\ref{sec:potentials} on the linear part of the Maris-Tandy potential.
The corresponding Chew-Frautschi plots are displayed in Fig.~\ref{fig:chew-frautschi}.
For all of these plots, a linear fit of the form $M^2(J) = M_0^2 + \beta J$ has been performed.
The resulting best fit parameters can be found in Tab.~\ref{tab:regge-fits}.
The calculated masses are mostly in agreement with the linear fits within the numerical uncertainty.

While in the past similar results have been produced within the same framework~\cite{Fischer:2014xha}, there was no
obvious reason as to why Regge behaviour should arise from the truncation and the used effective running coupling.
The emergence of a linearly rising term in the potential extracted from the running coupling provides some explanation
of why Regge behaviour can be observed independent of quark flavour content.
Notably, the heavy quark results are best described by a linear fit.
This is in agreement with the notion that a non-relativistic description and an explanation in terms of an interaction potential should become valid only in the heavy quark sector.

\section{Conclusion}\label{sec:conclusion}

In this work we discussed first exploratory results from functional methods for quark-antiquark states with total angular momentum 
$J=4,5$ in the light, open and hidden strange, hidden charm and hidden bottom flavour sectors of QCD. To this end,
we employed an approach using the relativistic Dyson-Schwinger and Bethe-Salpeter equations of QCD in a simple
rainbow-ladder type truncation. This truncation was suggested to work well for a sequence of states on the Regge 
trajectory with quantum numbers $J^{PC}=1^{--}, 2^{++}, 3^{--}, 4^{++}, 5^{--}$ in previous 
works~\cite{Fischer:2014xha,Fischer:2014cfa}. We confirm this notion in all cases where experimental results for 
the $J^{PC}=4^{++}$ are available and generate predictions for the ground state masses of states with 
$J^{PC}=4^{++}, 5^{--}$ in all flavour sectors. In general, we also observe that the rainbow-ladder truncation
gets better and better with increasing quark masses, confirming the notion of Ref.~\cite{Watson:2012ht} albeit
in a different gauge (Coulomb gauge) than the Landau gauge used in this work. We also verified that we obtain
straight lines in Chew-Frautschi plots on the 'good' Regge trajectory. This behaviour may be tied to the specific 
momentum dependence of our interaction model which results in a linear rising heavy-quark potential at intermediate 
distances. Let us emphasize again that this behaviour is accidental in the sense that the interaction model has not 
been chosen with this property in mind \cite{Maris:1997tm,Maris:1997hd,Maris:1999nt}. Nor does it in any intentional 
way reflect the underlying physics of an asymptotic string tension associated, e.g., with the physics of topologically 
non-trivial field configurations \cite{Greensite:2011zz}. But it is, to our mind, a highly amusing property of an
interaction model that has been successfully applied to many aspects of hadron physics in the past decades, see
e.g. \cite{Maris:2003vk,Cloet:2013jya,Eichmann:2016yit} for overviews.

\vspace*{3mm}
{\bf Acknowledgments}\\
	This work was supported by the DFG under grant numbers FI 970/11-2 and HU 2176/3-1, 
	the graduate school HGS-HIRe the 'Studienstiftung des deutschen Volkes' and 
	the GSI Helmholtzzentrum f\"ur Schwerionenforschung. 
	This work contributes to the aims of the U.S. Department of Energy ExoHad Topical Collaboration, contract DE-SC0023598. 
	We acknowledge computational resources provided by the HPC Core Facility and the HRZ of the Justus-Liebig-Universit\"at Gie\ss en.

\appendix

\section{Extrapolation of eigenvalue curves}\label{sec:app:extrapol}

\begin{figure}
	\includegraphics[width=\linewidth]{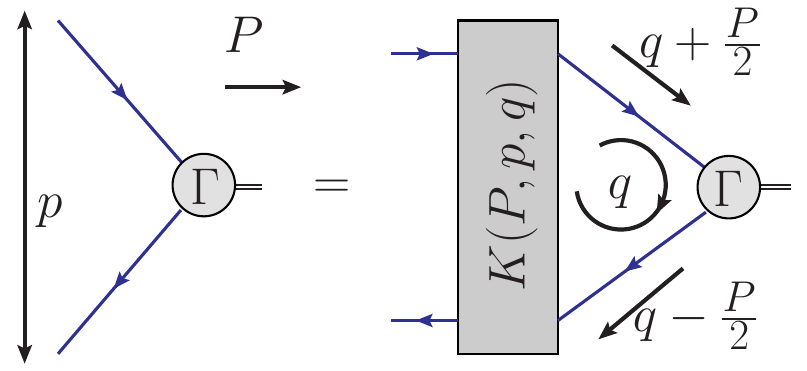}
	\caption{
		Momentum routing in the BSE. The total momentum $P$ is split between the two internal legs.
	}
	\label{fig:routing}
\end{figure}

When solving the meson BSE, one needs to evaluate the quark propagator's dressing functions at momenta
\begin{align}\label{eq:qpm}
	q_\pm^2 &= \left(q\pm\frac{P}{2}\right)^2= \nonumber\\
	 &=q^2 + \frac14 P^2 \pm q\cdot P = |q|^2 - \frac{1}{4}m^2 \pm i |q| m \cos \phi,
\end{align}
see Fig.~\ref{fig:routing},
where $\phi$ is the angle between $q$ and $P$.
This expression is valid in the center of mass frame of the meson, where one can write the total on-shell momentum of
the meson as $P=(\vec 0, i\,m)$ with $m\in \mathds{R}$.
This necessitates an evaluation of the quark propagator within a parabolic region in the complex plane.
However, the rainbow-ladder truncation is known to produce complex conjugate poles and other non-analytic structures in
the complex quark propagator~\cite{Alkofer:2003jj,Dorkin:2013rsa,Windisch:2016iud}.
This gives rise to a limit $P_0^2$ on how small $P^2$ can get for a direct evaluation of the Bethe-Salpeter kernel matrix.
Thus, for $P^2 < P_0^2$, we extrapolate the eigenvalue curve of the Bethe-Salpeter matrix as a function $\lambda = \lambda(P^2)$ and
determine the value of $P^2$ where it is equal to one.

An approach, which has proven to work quite well for this purpose, is to calculate a number of eigenvalues on a grid of
different values of $P^2$ and use a rational function with coefficients determined by the Schlessinger method~\cite{Schlessinger:1968spm}.
The idea behind this method is to write the rational function as a continued fraction
\begin{equation}\label{eq:schlessinger-function}
	C_n(x)=\frac{f_1}{1+\frac{a_1(x-x_1)}{1+\frac{a_2(x-x_2)}{\vdots a_{n-1}(x-x_{n-1})}}},
\end{equation}
where $x_1,\dots, x_n\equiv P_1^2,\dots, P_n^2$ are the points in the $P^2$ grid and $f_i=f(x_i)\equiv\lambda(P_i^2)$ are
the explicitly calculated eigenvalues.
The coefficients $a_i$ can be calculated by initializing $a_1 = \frac{f_1/f_2 - 1}{x_2-x_1}$ and then using the
iterative formula
\begin{equation}\label{eq:schlessinger-coefficients}
	a_i=\frac{1}{x_i-x_{i+1}}\left( 1+ \frac{a_{i-1}(x_{i+1}-x_{i-1})}{1+
		\frac{a_{i-2}(x_{i+1}-x_{i-2})}{\vdots \frac{a_1(x_{i+1}-x_1)}{1-f_1/f_{i+1}}}} \right).
\end{equation}

\begin{figure}
	\includegraphics[width=\linewidth]{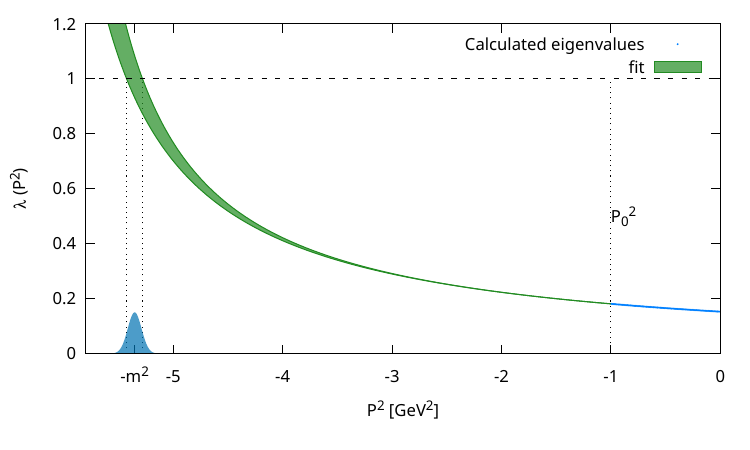}
	\caption{
		Schematic depiction of the extrapolation of the eigenvalue curve.
	}
	\label{fig:resampling}
\end{figure}

We estimate the numerical error introduced with this method via a resampling of the data points $(x_i, f_i)$.
This is done by taking a random subset of size $n<N$ of the original data and performing a Schlessinger extrapolation on that
subset to get an estimate of $P_j^2=-m_j^2$ where $C_{n,j}(x_j\equiv P_j^2) = 1$.
This sampling is repeated $N_R$ times and the mean and standard deviation of the resulting set of values for
$P^2=-m^2$ are taken as the final result for the meson mass and its numerical error.
We used the values $N = 125$ and $N_R = 1000$.
This is depicted schematically in Fig.~\ref{fig:resampling}.

The furthest momentum for which the eigenvalue has been calculated explicitly $P_0^2$ is given in Tab.~\ref{tab:extrapolation-momenta}.
Additionally, if no solution can be found for $P^2 > P^2_{\max{}}$, we discard that sample.
The values for $P^2_{\max{}}$ are also listed in Tab.~\ref{tab:extrapolation-momenta}.

\begin{table*}
	\begin{center}
		\begin{tabular}{c|cccc}
			\hline
			\hline
			           & up/down          & strange          & charm       & bottom \\
			\hline
			Momentum interval    & $[-1, -10^{-2}]$ & $[-1, -10^{-2}]$ & $[-12, -6]$ & $[-100, -72]$ \\
			$P^2_{\max{}}$ & $-30$            & $-30$            & $-36$       & $-144$   \\
			\hline
			\hline
		\end{tabular}
	\end{center}
	\caption{Momentum intervals, for which eigenvalues were calculated explicitly and threshold momentum for when samples were discarded (given in units of GeV$^2$).}\label{tab:extrapolation-momenta}
\end{table*}

\section{Tensor bases for mesons of arbitrary angular momentum}\label{sec:app:bases}

When solving the Bethe-Salpeter equation for mesons, their quantum numbers are reflected in their Bethe-Salpeter
amplitudes.
Thus, if one wants to solve the BSE for arbitrary quantum numbers, one needs a way to construct a tensor basis with the
corresponding quantum numbers.
While for states with $J=0$ and $J=1$ these bases can be constructed rather easily, higher angular momenta are best
handled with a systematic approach, which is sketched in~\cite{Eichmann:2016yit,Sanchis-Alepuz:2017jjd} and will be explained in detail
here.

\subsection{Constraints on the Bethe-Salpeter amplitude}\label{subsec:bsa-constraints}

In order to represent a unique physical meson state, a Bethe-Salpeter amplitude $\Gamma^{\muonetoJ}(p,P)$ has to satisfy a
number of constraints~\cite{Fierz:1939zz}:
\begin{enumerate}
	\item The amplitude has to be transverse with respect to the meson's momentum, i.e.
	$\Gamma^{\muonetoJ}(p,P)P_{\mu_1}=0$.
	\item The number of Lorentz indices of the amplitude has to equal the total angular momentum $J$ of the state.
	\item The amplitude has to be symmetric in its Lorentz indices, i.e.
	$\Gamma^{\muonetoJ}(p,P)=\Gamma^{\Pi(\muonetoJ)}(p,P)$ where $\Pi\in\text{S}_J$ is an arbitrary permutation of
	the Lorentz indices.
	\item The Lorentz trace $\Gamma^{\mu_1\phantom{\mu_2}\dots\mu_J}_{\phantom{\mu_1}\mu_1}$ of the amplitude has to be zero.
	\item It has to have the correct behaviour under a parity transformation.
\end{enumerate}
Note that due to the third requirement the amplitude is in fact transverse and traceless in all of its Lorentz indices.
Our goal is now to find a set of linearly independent tensors $\{\tau_i^{\muonetoJ}\}$ which form a basis of the vector
space of the amplitude such that we can write it as
\begin{equation}\label{eq:bsa-basis-general}
	\Gamma^{\muonetoJ}(p,P)=\sum_{i=1}^{n}f_i(p^2,P^2,p\cdot P)\tau_i^{\muonetoJ}(p,P),
\end{equation}
where the dressing functions $f_i$ only depend on Lorentz invariant combinations of $p$ and $P$.
It is then sufficient to construct the $\tau_i$ in a way to fulfill the aforementioned constraints for the entire
amplitude to fulfill them as well.
Additionally, in practice it is extremely beneficial if the chosen basis is orthonormal, as this makes projection onto
the dressing functions trivial and speeds up numerical computations.
We will now go over each criterion and outline a procedure to construct a basis for arbitrary values of $J$.

\subsection{Constructing a basis}\label{subsec:constructing-a-basis}

\subsubsection{Transversality}

The canonical choice of building blocks for the tensor basis are the unit matrix $\mathds{1}$, the relative and absolute
four-momenta $p^\mu$ and $P^\mu$ as well as their contractions with the gamma matrix four vector $\gamma^\mu$, i.e.
$\slashed p = \gamma^\mu p^\mu$ and $\slashed P = \gamma^\mu P^{\mu}$, and the gamma matrix four vector itself.
These are, however, neither normalized nor transverse to the total momentum $P^\mu$, and it is thus more convenient to
work with the normalized and transverse versions of these quantities, i. e. 
\begin{align}\label{eq:transverse-quantities}
	\hat{P}^\mu &= \frac{P^\mu}{P}, \\
	p_\perp^\mu &= p^\mu - (p\cdot \hat{P})\,\hat{P}^\mu, \\
	n^\mu &= \widehat{p_\perp}^\mu = \frac{\hat{p}^\mu - z\,\hat{P}^\mu}{\sqrt{1-z^2}}\,,
\end{align}
where $z=\hat{p}\cdot \hat{P}$.
Note that this allows for a choice of coordinates where $\hat{P}^\mu = (0, 0, 0, 1)$ and $n^\mu = (0, 0, 1, 0)$.
These quantities have the convenient properties that $n^2 = \hat{P}^2 = 1$ and $n\cdot \hat{P} = 0$.
Along with these redefinitions one can define transverse gamma matrices with respect to both $n$ and $\hat{P}$ via
\begin{align}\label{eq:transverse-gammas}
\gamma^\mu_\perp &= T^{\mu\nu}_P\,\gamma^\nu = \gamma^\mu - \hat{P}^\mu \hat{\slashed{P}} \\
\gamma^\mu_{\perp\perp} &= \gamma^\mu - n^\mu \slashed{n} - \hat{P}^\mu \hat{\slashed{P}}\,,
\end{align}
where $T_P^{\mu\nu} = \delta^{\mu\nu} - \hat{P}^\mu\hat{P}^\nu$ is the transverse projector with respect to $P^\mu$.
These transverse gamma matrices anticommute with both $\slashed n$ and $\hat{\slashed P}$.

\subsubsection{Angular momentum}

With these building blocks at hand, one can construct the basis of a state with total angular momentum $J$ as
\begin{equation}\label{eq:meson-basis-1}
	\begin{split}
		J=0: &\quad  \{ \, \mathds{1}, \; \hat{\slashed{P}} \,\} \, \{ \, \mathds{1}, \;\, \slashed{n} \, \}\,, \\
		J>0: &\quad  \{ \, \mathds{1}, \; \hat{\slashed{P}} \,\} \, \{ \, \mathds{1}, \;\, \slashed{n} \, \} \, \{ \, N^{\muonetoJ}, \;\; M^{\muonetoJ} \} \,.
	\end{split}
\end{equation}
The tensor $N^{\muonetoJ}$ can contain all possible combinations of $n^{\mu}$ and $T_P^{\mu\nu}$ while being
transverse to the total momentum $P^\mu$:
\begin{multline}\label{eq:n-tensor-even}
	n^{\mu_1}n^{\mu_2}\cdots n^{\mu_J}, \quad T_P^{\mu_1\mu_2}n^{\mu_3}\cdots n^{\mu_J}, \quad\dots ,\\ T_P^{\mu_1\mu_2}T_P^{\mu_3\mu_4}\cdots T_P^{\mu_{J-1}\mu_J}.
\end{multline}
Note that for odd values of $J$ there is one more term containing $(J-1)/2$ copies of $T_P$ and a single $n$.
The tensor $M^{\muonetoJ}$, on the other hand, contains in addition a single factor of $\gamma^\mu_{\perp\perp}$ in every term:
\begin{multline}\label{eq:m-tensor-even}
	\gamma_{\perp\perp}^{\mu_1}n^{\mu_2}\cdots n^{\mu_J}, \quad \gamma_{\perp\perp}^{\mu_1}T_P^{\mu_2\mu_3}n^{\mu_4}\cdots n^{\mu_J}, \quad\dots ,\\
	 \gamma_{\perp\perp}^{\mu_1}T_P^{\mu_2\mu_3}T_P^{\mu_4\mu_5}\dots T_P^{\mu_{J-2}\mu_{J-1}}n^{\mu_J}.
\end{multline}
The last term is only present for even values of $J$.

\begin{table*}
	\centering
	\begin{tabular}{c|c|c}
		$i$ & $\tau_i$                                     & $\bar{\tau}_i$ \\
		\hline
		1   & $\frac{1}{2} \mathds{1}$                     & $\tau_1$       \\
		2   & $\frac{1}{2} i \hat{\slashed{P}}$            & $-\tau_2$      \\
		3   & $\frac{1}{2} i \slashed{n}$                  & $-\tau_3$      \\
		4   & $\frac{1}{2} i \hat{\slashed{P}}\slashed{n}$ & $\tau_4$
	\end{tabular}\hspace*{1cm}
	\begin{tabular}{c|c|c}
		$i$ & $\tau_i$                                              & $\bar{\tau}_i$ \\
		\hline
		1   & $\frac{1}{2} \gamma_5$                                & $\tau_1$       \\
		2   & $\frac{1}{2} \gamma_5 \hat{\slashed{P}}$              & $-\tau_2$      \\
		3   & $\frac{1}{2} \gamma_5 \slashed{n}$                    & $-\tau_3$      \\
		4   & $\frac{1}{2} i \gamma_5 \hat{\slashed{P}}\slashed{n}$ & $\tau_4$
	\end{tabular}\hspace*{1cm}
	\begin{tabular}{c|c|c}
		$i$ & $\tau_i$                                                                  & $\bar{\tau}_i$ \\
		\hline
		1   & $\frac{1}{2}      i \mathds{1} n^\mu                                    $ & $-\tau_1$      \\
		2   & $\frac{1}{2}        \hat{\slashed{P}} n^\mu                             $ & $\tau_2$       \\
		3   & $\frac{1}{2}        \slashed{n} n^\mu                                   $ & $\tau_3$       \\
		4   & $\frac{1}{2}        \hat{\slashed{P}}\slashed{n} n^\mu                  $ & $-\tau_4$      \\
		5   & $\frac{\sqrt2}{4}   \gamma_{\perp\perp}^\mu                             $ & $\tau_5$       \\
		6   & $\frac{\sqrt2}{4}   \hat{\slashed{P}} \gamma_{\perp\perp}^\mu           $ & $-\tau_6$      \\
		7   & $\frac{\sqrt2}{4}   \slashed{n} \gamma_{\perp\perp}^\mu                 $ & $-\tau_7$      \\
		8   & $\frac{\sqrt2}{4} i \hat{\slashed{P}}\slashed{n} \gamma_{\perp\perp}^\mu$ & $\tau_8$
	\end{tabular}\hspace*{1cm}
	\begin{tabular}{c|c|c}
		$i$ & $\tau_i$                                                                          & $\bar{\tau}_i$ \\
		\hline
		1   & $\frac{1}{2}   \gamma_5 n^\mu                                                   $ & $\tau_1$       \\
		2   & $\frac{1}{2}   \gamma_5\hat{\slashed{P}} n^\mu                                  $ & $-\tau_2$      \\
		3   & $\frac{1}{2}   \gamma_5\slashed{n} n^\mu                                        $ & $-\tau_3$      \\
		4   & $\frac{1}{2} i \gamma_5\hat{\slashed{P}}\slashed{n} n^\mu                       $ & $\tau_4$       \\
		5   & $\frac{\sqrt2}{4}   \gamma_5\gamma_{\perp\perp}^\mu                             $ & $-\tau_5$      \\
		6   & $\frac{\sqrt2}{4} i \gamma_5\hat{\slashed{P}} \gamma_{\perp\perp}^\mu           $ & $\tau_6$       \\
		7   & $\frac{\sqrt2}{4} i \gamma_5\slashed{n} \gamma_{\perp\perp}^\mu                 $ & $\tau_7$       \\
		8   & $\frac{\sqrt2}{4} i \gamma_5\hat{\slashed{P}}\slashed{n} \gamma_{\perp\perp}^\mu$ & $-\tau_8$
	\end{tabular}
	\caption{Tensor structures for mesons with quantum numbers (from left to right) $J^P=0^{+}$, $J^P=0^{-}$, $J^P=1^{-}$, $J^P=1^{+}$.}
	\label{tab:bases1}
	\begin{tabular}{c|c|c}
		$i$ & $\tau_i$                                                                                                                       & $\bar{\tau}_i$ \\
		\hline
		1   & $\frac{\sqrt6}{4} i \mathds{1} \left( n^\mu n^\nu - \frac13 T_P^{\mu\nu} \right)                                             $ & $-\tau_1$      \\
		2   & $\frac{\sqrt6}{4}   \hat{\slashed{P}} \left( n^\mu n^\nu - \frac13 T_P^{\mu\nu} \right)                                      $ & $\tau_2$       \\
		3   & $\frac{\sqrt6}{4}   \slashed{n} \left( n^\mu n^\nu - \frac13 T_P^{\mu\nu} \right)                                            $ & $\tau_3$       \\
		4   & $\frac{\sqrt6}{4}   \hat{\slashed{P}}\slashed{n} \left( n^\mu n^\nu - \frac13 T_P^{\mu\nu} \right)                           $ & $-\tau_4$      \\
		5   & $\frac{1}{4}        \left( \gamma_{\perp\perp}^\mu n^\nu + n^\mu \gamma_{\perp\perp}^\nu \right)                             $ & $\tau_5$       \\
		6   & $\frac{1}{4}        \hat{\slashed{P}}\left( \gamma_{\perp\perp}^\mu n^\nu + n^\mu \gamma_{\perp\perp}^\nu \right)            $ & $-\tau_6$      \\
		7   & $\frac{1}{4}        \slashed{n} \left( \gamma_{\perp\perp}^\mu n^\nu + n^\mu \gamma_{\perp\perp}^\nu \right)                 $ & $-\tau_7$      \\
		8   & $\frac{1}{4}      i \hat{\slashed{P}}\slashed{n} \left( \gamma_{\perp\perp}^\mu n^\nu + n^\mu \gamma_{\perp\perp}^\nu \right)$ & $\tau_8$
	\end{tabular}\hspace*{1cm}
	\begin{tabular}{c|c|c}
		$i$ & $\tau_i$                                                                                                                                & $\bar{\tau}_i$ \\
		\hline
		1   & $\frac{\sqrt6}{4}   \gamma_5 \left( n^\mu n^\nu - \frac13 T_P^{\mu\nu} \right)                                                        $ & $\tau_1$       \\
		2   & $\frac{\sqrt6}{4}   \gamma_5 \hat{\slashed{P}} \left( n^\mu n^\nu - \frac13 T_P^{\mu\nu} \right)                                      $ & $-\tau_2$      \\
		3   & $\frac{\sqrt6}{4}   \gamma_5 \slashed{n} \left( n^\mu n^\nu - \frac13 T_P^{\mu\nu} \right)                                            $ & $-\tau_3$      \\
		4   & $\frac{\sqrt6}{4} i \gamma_5 \hat{\slashed{P}}\slashed{n} \left( n^\mu n^\nu - \frac13 T_P^{\mu\nu} \right)                           $ & $\tau_4$       \\
		5   & $\frac{1}{4}        \gamma_5 \left( \gamma_{\perp\perp}^\mu n^\nu + n^\mu \gamma_{\perp\perp}^\nu \right)                             $ & $-\tau_5$      \\
		6   & $\frac{1}{4}      i \gamma_5 \hat{\slashed{P}}\left( \gamma_{\perp\perp}^\mu n^\nu + n^\mu \gamma_{\perp\perp}^\nu \right)            $ & $\tau_6$       \\
		7   & $\frac{1}{4}      i \gamma_5 \slashed{n} \left( \gamma_{\perp\perp}^\mu n^\nu + n^\mu \gamma_{\perp\perp}^\nu \right)                 $ & $\tau_7$       \\
		8   & $\frac{1}{4}      i \gamma_5 \hat{\slashed{P}}\slashed{n} \left( \gamma_{\perp\perp}^\mu n^\nu + n^\mu \gamma_{\perp\perp}^\nu \right)$ & $-\tau_8$
	\end{tabular}
	\caption{Tensor structures for mesons with quantum numbers (from left to right) $J^P=2^{+}$, $J^P=2^{-}$.}
	\label{tab:bases2}
	\begin{tabular}{c|c|c}
		$i$ & $\tau_i$                                                                                                                                                               & $\bar{\tau}_i$ \\
		\hline
		1   & $\frac{\sqrt{10}}{4} i  \mathds{1} \left( n^\mu n^\nu n^\rho - \frac15 T_P^{[\mu\nu} n^{\rho]} \right)                                                               $ & $-\tau_1$      \\
		2   & $\frac{\sqrt{10}}{4}    \hat{\slashed{P}} \left( n^\mu n^\nu n^\rho - \frac15 T_P^{[\mu\nu} n^{\rho]} \right)                                                        $ & $\tau_2$       \\
		3   & $\frac{\sqrt{10}}{4}    \slashed{n}\left( n^\mu n^\nu n^\rho - \frac15 T_P^{[\mu\nu} n^{\rho]} \right)                                                               $ & $\tau_3$       \\
		4   & $\frac{\sqrt{10}}{4}    \hat{\slashed{P}}\slashed{n} \left( n^\mu n^\nu n^\rho - \frac15 T_P^{[\mu\nu} n^{\rho]} \right)                                             $ & $-\tau_4$      \\
		5   & $\frac{\sqrt{30}}{24}   \left( \gamma_{\perp\perp}^{[\mu} n^\nu n^{\rho]} - \frac{1}{5} \gamma_{\perp\perp}^{[\mu}T_P^{\nu\rho]} \right)                             $ & $\tau_5$       \\
		6   & $\frac{\sqrt{30}}{24}   \hat{\slashed{P}}\left(\gamma_{\perp\perp}^{[\mu} n^\nu n^{\rho]} - \frac{1}{5} \gamma_{\perp\perp}^{[\mu}T_P^{\nu\rho]} \right)             $ & $-\tau_6$      \\
		7   & $\frac{\sqrt{30}}{24}   \slashed{n} \left( \gamma_{\perp\perp}^{[\mu} n^\nu n^{\rho]} - \frac{1}{5} \gamma_{\perp\perp}^{[\mu}T_P^{\nu\rho]} \right)                 $ & $-\tau_7$      \\
		8   & $\frac{\sqrt{30}}{24} i \hat{\slashed{P}}\slashed{n} \left( \gamma_{\perp\perp}^{[\mu} n^\nu n^{\rho]} - \frac{1}{5} \gamma_{\perp\perp}^{[\mu}T_P^{\nu\rho]} \right)$ & $\tau_8$
	\end{tabular}\hspace*{1cm}
	\begin{tabular}{c|c|c}
		$i$ & $\tau_i$                                                                                                                                                                        & $\bar{\tau}_i$ \\
		\hline
		1   & $\frac{\sqrt{10}}{4}    \gamma_5 \mathds{1} \left( n^\mu n^\nu n^\rho - \frac15 T_P^{[\mu\nu} n^{\rho]} \right)                                                               $ & $\tau_1$       \\
		2   & $\frac{\sqrt{10}}{4}    \gamma_5 \hat{\slashed{P}} \left( n^\mu n^\nu n^\rho - \frac15 T_P^{[\mu\nu} n^{\rho]} \right)                                                        $ & $-\tau_2$      \\
		3   & $\frac{\sqrt{10}}{4}    \gamma_5 \slashed{n}\left( n^\mu n^\nu n^\rho - \frac15 T_P^{[\mu\nu} n^{\rho]} \right)                                                               $ & $-\tau_3$      \\
		4   & $\frac{\sqrt{10}}{4} i  \gamma_5 \hat{\slashed{P}}\slashed{n} \left( n^\mu n^\nu n^\rho - \frac15 T_P^{[\mu\nu} n^{\rho]} \right)                                             $ & $\tau_4$       \\
		5   & $\frac{\sqrt{30}}{24}   \gamma_5 \left( \gamma_{\perp\perp}^{[\mu} n^\nu n^{\rho]} - \frac{1}{5} \gamma_{\perp\perp}^{[\mu}T_P^{\nu\rho]} \right)                             $ & $-\tau_5$      \\
		6   & $\frac{\sqrt{30}}{24} i \gamma_5 \hat{\slashed{P}}\left(\gamma_{\perp\perp}^{[\mu} n^\nu n^{\rho]} - \frac{1}{5} \gamma_{\perp\perp}^{[\mu}T_P^{\nu\rho]} \right)             $ & $\tau_6$       \\
		7   & $\frac{\sqrt{30}}{24} i \gamma_5 \slashed{n} \left( \gamma_{\perp\perp}^{[\mu} n^\nu n^{\rho]} - \frac{1}{5} \gamma_{\perp\perp}^{[\mu}T_P^{\nu\rho]} \right)                 $ & $\tau_7$       \\
		8   & $\frac{\sqrt{30}}{24} i \gamma_5 \hat{\slashed{P}}\slashed{n} \left( \gamma_{\perp\perp}^{[\mu} n^\nu n^{\rho]} - \frac{1}{5} \gamma_{\perp\perp}^{[\mu}T_P^{\nu\rho]} \right)$ & $-\tau_8$
	\end{tabular}
	\caption{Tensor structures for mesons with quantum numbers (from left to right) $J^P=3^{-}$, $J^P=3^{+}$.}	
	\label{tab:bases3}
	\end{table*}

\subsubsection{Symmetry}

To make the tensor basis symmetric in all its Lorentz indices, we sum up all permutations of indices for each term.
This leaves us with the following general form of $N^{\muonetoJ}$ and $M^{\muonetoJ}$ for even $J$
\begin{equation}\label{eq:n-m-general}
	\begin{split}
		N^{\muonetoJ} =& N_{J,0} ( n^{\mu_1}n^{\mu_2}\cdots n^{\mu_J} \\
		&+ N_{J,1} T_P^{[\mu_1\mu_2}n^{\mu_3}\cdots n^{\mu_J]} \\
		&+ \dots   + N_{J,J/2}T_P^{[\mu_1\mu_2}T_P^{\mu_3\mu_4}\cdots T_P^{\mu_{J-1}\mu_J]} )\\
		M^{\muonetoJ} =& M_{J,0} ( \gamma_{\perp\perp}^{[\mu_1}n^{\mu_2}\cdots n^{\mu_J]} \\
		&+ M_{J,1} \gamma_{\perp\perp}^{[\mu_1}T_P^{\mu_2\mu_3}n^{\mu_4}\cdots n^{\mu_J]} + \dots   \\
		+M_{J,(J-2)/2} &\gamma_{\perp\perp}^{[\mu_1}T_P^{\mu_2\mu_3}T_P^{\mu_4\mu_5}\dots T_P^{\mu_{J-2}\mu_{J-1}}n^{\mu_J]} ).
	\end{split}
\end{equation}
The brackets around the Lorentz indices indicate a sum over all unique permutations of the indices and the coefficients
$N_{J,i}$ and $M_{J,i}$ have to be chosen such that the final basis is traceless and normalized.
In practice, this is a bottleneck for going to arbitrarily large angular momenta since the number of terms needed for
symmetrization grows very rapidly for increasing $J$.
To see this, we can find a lower bound for the number of terms needed by just considering even $J$ and only the term
\begin{equation}\label{eq:lower-bound1}
	T_P^{[\mu_1\mu_2}T_P^{\mu_3\mu_4}\cdots T_P^{\mu_{J-1}\mu_J]} .
\end{equation}
There is a total of $J!$ possible permutations of indices in that term alone, but due to the symmetry of $T_P^{\mu\nu}$ the
number of unique permutations of this term is $J!/2^J$, which still grows faster than exponentially for sufficiently large $J$.

\begin{table*}
	\centering
	\begin{tabular}{c|c|c}
			$i$ & $\tau_i$                                                                                                                                                                                    & $\bar{\tau}_i$ \\
			\hline
			1   & $\frac{\sqrt{70}}{8} i  \mathds{1}                   \left( n^\mu n^\nu n^\rho n^\sigma - \frac17 T_P^{[\mu\nu} n^{\rho} n^{\sigma]} + \frac{1}{35} T_P^{[\mu\nu}T_P^{\rho\sigma]} \right)$ & $-\tau_1$      \\
			2   & $\frac{\sqrt{70}}{8}    \hat{\slashed{P}}            \left( n^\mu n^\nu n^\rho n^\sigma - \frac17 T_P^{[\mu\nu} n^{\rho} n^{\sigma]} + \frac{1}{35} T_P^{[\mu\nu}T_P^{\rho\sigma]} \right)$ & $\tau_2$       \\
			3   & $\frac{\sqrt{70}}{8}    \slashed{n}                  \left( n^\mu n^\nu n^\rho n^\sigma - \frac17 T_P^{[\mu\nu} n^{\rho} n^{\sigma]} + \frac{1}{35} T_P^{[\mu\nu}T_P^{\rho\sigma]} \right)$ & $\tau_3$       \\
			4   & $\frac{\sqrt{70}}{8}    \hat{\slashed{P}}\slashed{n} \left( n^\mu n^\nu n^\rho n^\sigma - \frac17 T_P^{[\mu\nu} n^{\rho} n^{\sigma]} + \frac{1}{35} T_P^{[\mu\nu}T_P^{\rho\sigma]} \right)$ & $-\tau_4$      \\
			5   & $\frac{\sqrt{56}}{32}                                \left( \gamma_{\perp\perp}^{[\mu} n^\nu n^\rho n^{\sigma]} - \frac{1}{7} \gamma_{\perp\perp}^{[\mu}T_P^{\nu\rho} n^{\sigma]}  \right)$ & $\tau_5$       \\
			6   & $\frac{\sqrt{56}}{32}   \hat{\slashed{P}}            \left( \gamma_{\perp\perp}^{[\mu} n^\nu n^\rho n^{\sigma]} - \frac{1}{7} \gamma_{\perp\perp}^{[\mu}T_P^{\nu\rho} n^{\sigma]}  \right)$ & $-\tau_6$      \\
			7   & $\frac{\sqrt{56}}{32}   \slashed{n}                  \left( \gamma_{\perp\perp}^{[\mu} n^\nu n^\rho n^{\sigma]} - \frac{1}{7} \gamma_{\perp\perp}^{[\mu}T_P^{\nu\rho} n^{\sigma]}  \right)$ & $-\tau_7$      \\
			8   & $\frac{\sqrt{56}}{32} i \hat{\slashed{P}}\slashed{n} \left( \gamma_{\perp\perp}^{[\mu} n^\nu n^\rho n^{\sigma]} - \frac{1}{7} \gamma_{\perp\perp}^{[\mu}T_P^{\nu\rho} n^{\sigma]}  \right)$ & $\tau_8$
		\end{tabular}\hspace*{0cm}
	\begin{tabular}{c|c|c}
			$i$ & $\tau_i$                                                                                                                                                                                             & $\bar{\tau}_i$ \\
			\hline
			1   & $\frac{\sqrt{70}}{8}    \gamma_5 \mathds{1}                   \left( n^\mu n^\nu n^\rho n^\sigma - \frac17 T_P^{[\mu\nu} n^{\rho} n^{\sigma]} + \frac{1}{35} T_P^{[\mu\nu}T_P^{\rho\sigma]} \right)$ & $\tau_1$       \\
			2   & $\frac{\sqrt{70}}{8}    \gamma_5 \hat{\slashed{P}}            \left( n^\mu n^\nu n^\rho n^\sigma - \frac17 T_P^{[\mu\nu} n^{\rho} n^{\sigma]} + \frac{1}{35} T_P^{[\mu\nu}T_P^{\rho\sigma]} \right)$ & $-\tau_2$      \\
			3   & $\frac{\sqrt{70}}{8}    \gamma_5 \slashed{n}                  \left( n^\mu n^\nu n^\rho n^\sigma - \frac17 T_P^{[\mu\nu} n^{\rho} n^{\sigma]} + \frac{1}{35} T_P^{[\mu\nu}T_P^{\rho\sigma]} \right)$ & $-\tau_3$      \\
			4   & $\frac{\sqrt{70}}{8} i  \gamma_5 \hat{\slashed{P}}\slashed{n} \left( n^\mu n^\nu n^\rho n^\sigma - \frac17 T_P^{[\mu\nu} n^{\rho} n^{\sigma]} + \frac{1}{35} T_P^{[\mu\nu}T_P^{\rho\sigma]} \right)$ & $\tau_4$       \\
			5   & $\frac{\sqrt{56}}{32}   \gamma_5                              \left( \gamma_{\perp\perp}^{[\mu} n^\nu n^\rho n^{\sigma]} - \frac{1}{7} \gamma_{\perp\perp}^{[\mu}T_P^{\nu\rho} n^{\sigma]}  \right)$ & $-\tau_5$      \\
			6   & $\frac{\sqrt{56}}{32} i \gamma_5 \hat{\slashed{P}}            \left( \gamma_{\perp\perp}^{[\mu} n^\nu n^\rho n^{\sigma]} - \frac{1}{7} \gamma_{\perp\perp}^{[\mu}T_P^{\nu\rho} n^{\sigma]}  \right)$ & $\tau_6$       \\
			7   & $\frac{\sqrt{56}}{32} i \gamma_5 \slashed{n}                  \left( \gamma_{\perp\perp}^{[\mu} n^\nu n^\rho n^{\sigma]} - \frac{1}{7} \gamma_{\perp\perp}^{[\mu}T_P^{\nu\rho} n^{\sigma]}  \right)$ & $\tau_7$       \\
			8   & $\frac{\sqrt{56}}{32} i \gamma_5 \hat{\slashed{P}}\slashed{n} \left( \gamma_{\perp\perp}^{[\mu} n^\nu n^\rho n^{\sigma]} - \frac{1}{7} \gamma_{\perp\perp}^{[\mu}T_P^{\nu\rho} n^{\sigma]}  \right)$ & $-\tau_8$
		\end{tabular}
\caption{Tensor structures for mesons with quantum numbers (from left to right) $J^P=4^{+}$, $J^P=4^{-}$.}
	\label{tab:bases4}
	\begin{tabular}{c|c|c}
			$i$ & $\tau_i$                                                                                                                                                                                                                                                                                     & $\bar{\tau}_i$ \\
			\hline
			1   & $\frac{3\sqrt{14}}{8} i  \mathds{1}                   \left( n^\mu n^\nu n^\rho n^\sigma n^{\kappa} - \frac19 T_P^{[\mu\nu} n^{\rho} n^{\sigma} n^{\kappa]} + \frac{1}{63} T_P^{[\mu\nu}T_P^{\rho\sigma} n^{\kappa]}                                                                \right)$ & $-\tau_1$      \\
			2   & $\frac{3\sqrt{14}}{8}    \hat{\slashed{P}}            \left( n^\mu n^\nu n^\rho n^\sigma n^{\kappa} - \frac19 T_P^{[\mu\nu} n^{\rho} n^{\sigma} n^{\kappa]} + \frac{1}{63} T_P^{[\mu\nu}T_P^{\rho\sigma} n^{\kappa]}                                                                \right)$ & $\tau_2$       \\
			3   & $\frac{3\sqrt{14}}{8}    \slashed{n}                  \left( n^\mu n^\nu n^\rho n^\sigma n^{\kappa} - \frac19 T_P^{[\mu\nu} n^{\rho} n^{\sigma} n^{\kappa]} + \frac{1}{63} T_P^{[\mu\nu}T_P^{\rho\sigma} n^{\kappa]}                                                                \right)$ & $\tau_3$       \\
			4   & $\frac{3\sqrt{14}}{8}    \hat{\slashed{P}}\slashed{n} \left( n^\mu n^\nu n^\rho n^\sigma n^{\kappa} - \frac19 T_P^{[\mu\nu} n^{\rho} n^{\sigma} n^{\kappa]} + \frac{1}{63} T_P^{[\mu\nu}T_P^{\rho\sigma} n^{\kappa]}                                                                \right)$ & $-\tau_4$      \\
			5   & $\frac{\sqrt{105}}{40}                                \left( \gamma_{\perp\perp}^{[\mu} n^\nu n^\rho n^{\sigma} n^{\kappa]} - \frac{1}{9} \gamma_{\perp\perp}^{[\mu}T_P^{\nu\rho} n^{\sigma} n^{\kappa]} + \frac{1}{63} \gamma_{\perp\perp}^{[\mu}T_P^{\nu\rho} T_P^{\sigma\kappa]} \right)$ & $\tau_5$       \\
			6   & $\frac{\sqrt{105}}{40}   \hat{\slashed{P}}            \left( \gamma_{\perp\perp}^{[\mu} n^\nu n^\rho n^{\sigma} n^{\kappa]} - \frac{1}{9} \gamma_{\perp\perp}^{[\mu}T_P^{\nu\rho} n^{\sigma} n^{\kappa]} + \frac{1}{63} \gamma_{\perp\perp}^{[\mu}T_P^{\nu\rho} T_P^{\sigma\kappa]} \right)$ & $-\tau_6$      \\
			7   & $\frac{\sqrt{105}}{40}   \slashed{n}                  \left( \gamma_{\perp\perp}^{[\mu} n^\nu n^\rho n^{\sigma} n^{\kappa]} - \frac{1}{9} \gamma_{\perp\perp}^{[\mu}T_P^{\nu\rho} n^{\sigma} n^{\kappa]} + \frac{1}{63} \gamma_{\perp\perp}^{[\mu}T_P^{\nu\rho} T_P^{\sigma\kappa]} \right)$ & $-\tau_7$      \\
			8   & $\frac{\sqrt{105}}{40} i \hat{\slashed{P}}\slashed{n} \left( \gamma_{\perp\perp}^{[\mu} n^\nu n^\rho n^{\sigma} n^{\kappa]} - \frac{1}{9} \gamma_{\perp\perp}^{[\mu}T_P^{\nu\rho} n^{\sigma} n^{\kappa]} + \frac{1}{63} \gamma_{\perp\perp}^{[\mu}T_P^{\nu\rho} T_P^{\sigma\kappa]} \right)$ & $\tau_8$
		\end{tabular}\\
	\begin{tabular}{c|c|c}
			$i$ & $\tau_i$                                                                                                                                                                                                                                                                                              & $\bar{\tau}_i$ \\
			\hline
			1   & $\frac{3\sqrt{14}}{8}    \gamma_5 \mathds{1}                   \left( n^\mu n^\nu n^\rho n^\sigma n^{\kappa} - \frac19 T_P^{[\mu\nu} n^{\rho} n^{\sigma} n^{\kappa]} + \frac{1}{63} T_P^{[\mu\nu}T_P^{\rho\sigma} n^{\kappa]}                                                                \right)$ & $\tau_1$       \\
			2   & $\frac{3\sqrt{14}}{8}    \gamma_5 \hat{\slashed{P}}            \left( n^\mu n^\nu n^\rho n^\sigma n^{\kappa} - \frac19 T_P^{[\mu\nu} n^{\rho} n^{\sigma} n^{\kappa]} + \frac{1}{63} T_P^{[\mu\nu}T_P^{\rho\sigma} n^{\kappa]}                                                                \right)$ & $-\tau_2$      \\
			3   & $\frac{3\sqrt{14}}{8}    \gamma_5 \slashed{n}                  \left( n^\mu n^\nu n^\rho n^\sigma n^{\kappa} - \frac19 T_P^{[\mu\nu} n^{\rho} n^{\sigma} n^{\kappa]} + \frac{1}{63} T_P^{[\mu\nu}T_P^{\rho\sigma} n^{\kappa]}                                                                \right)$ & $-\tau_3$      \\
			4   & $\frac{3\sqrt{14}}{8} i  \gamma_5 \hat{\slashed{P}}\slashed{n} \left( n^\mu n^\nu n^\rho n^\sigma n^{\kappa} - \frac19 T_P^{[\mu\nu} n^{\rho} n^{\sigma} n^{\kappa]} + \frac{1}{63} T_P^{[\mu\nu}T_P^{\rho\sigma} n^{\kappa]}                                                                \right)$ & $\tau_4$       \\
			5   & $\frac{\sqrt{105}}{40}   \gamma_5                              \left( \gamma_{\perp\perp}^{[\mu} n^\nu n^\rho n^{\sigma} n^{\kappa]} - \frac{1}{9} \gamma_{\perp\perp}^{[\mu}T_P^{\nu\rho} n^{\sigma} n^{\kappa]} + \frac{1}{63} \gamma_{\perp\perp}^{[\mu}T_P^{\nu\rho} T_P^{\sigma\kappa]} \right)$ & $-\tau_5$      \\
			6   & $\frac{\sqrt{105}}{40} i \gamma_5 \hat{\slashed{P}}            \left( \gamma_{\perp\perp}^{[\mu} n^\nu n^\rho n^{\sigma} n^{\kappa]} - \frac{1}{9} \gamma_{\perp\perp}^{[\mu}T_P^{\nu\rho} n^{\sigma} n^{\kappa]} + \frac{1}{63} \gamma_{\perp\perp}^{[\mu}T_P^{\nu\rho} T_P^{\sigma\kappa]} \right)$ & $\tau_6$       \\
			7   & $\frac{\sqrt{105}}{40} i \gamma_5 \slashed{n}                  \left( \gamma_{\perp\perp}^{[\mu} n^\nu n^\rho n^{\sigma} n^{\kappa]} - \frac{1}{9} \gamma_{\perp\perp}^{[\mu}T_P^{\nu\rho} n^{\sigma} n^{\kappa]} + \frac{1}{63} \gamma_{\perp\perp}^{[\mu}T_P^{\nu\rho} T_P^{\sigma\kappa]} \right)$ & $\tau_7$       \\
			8   & $\frac{\sqrt{105}}{40} i \gamma_5 \hat{\slashed{P}}\slashed{n} \left( \gamma_{\perp\perp}^{[\mu} n^\nu n^\rho n^{\sigma} n^{\kappa]} - \frac{1}{9} \gamma_{\perp\perp}^{[\mu}T_P^{\nu\rho} n^{\sigma} n^{\kappa]} + \frac{1}{63} \gamma_{\perp\perp}^{[\mu}T_P^{\nu\rho} T_P^{\sigma\kappa]} \right)$ & $-\tau_8$
		\end{tabular}
\caption{Tensor structures for mesons with quantum numbers (from top to bottom) $J^P=5^{-}$, $J^P=5^{+}$.}
	\label{tab:bases5}
\end{table*}
\subsubsection{Tracelessness}

Since the expressions for $N^{\muonetoJ}$ and $M^{\muonetoJ}$ are constructed to be symmetric under permutation of Lorentz
indices, one can rather easily derive expressions for the coefficients $N_{J,i}$ and $M_{J,i}$ with $i\geq 1$ such that
their Lorentz trace vanishes.
A straightforward way of determining them is to evaluate the traces of $N$ and $M$ with respect to $\mu_1$ and
$\mu_2$ or any other pair of indices and tune them via comparison of coefficients to set the traces to zero.
Some helpful relations for this derivation are
\begin{align*}\label{eq:useful-relations1}
	n^{\mu}n_{\mu} =& 1,\quad T_P^{\mu\nu} n_{\nu} = n^{\mu},\quad T_{P\phantom{\nu}\nu}^{\phantom{P}\nu} = 3,\\
	\gamma_{\perp\perp}^{\mu}n_{\mu} =& 0,\quad T_P^{\mu\nu} T_{P\nu}^{\phantom{P\nu}\rho} = T_P^{\mu\rho},\quad T_P^{\mu\nu} \gamma_{\perp\perp\,\nu} = \gamma_{\perp\perp}^{\mu}.
\end{align*}
For example, in the case of $J=3$ one can construct the coefficients $N_{J,i}$ by first writing down all relevant terms
\begin{equation}\label{eq:example-j-3}
	\frac{1}{N_{3,0}} N^{\mu\nu\rho} = n^{\mu}n^{\nu}n^{\rho} + N_{3,1} \left( T_P^{\mu\nu} n^\rho + T_P^{\nu\rho} n^\mu + T_P^{\rho\mu} n^\nu \right).
\end{equation}
Taking the Lorentz trace with respect to $\mu$ and $\nu$ yields
\begin{align}\label{eq:example-j-3-trace}
	\frac{1}{N_{3,0}} N^{\mu\phantom{\mu}\rho}_{\phantom{\mu}\mu} &= n^{\mu}n_{\mu}n^{\rho} + N_{3,1} \left( T_{P\phantom{\mu}\mu}^{\phantom{P}\mu} n^\rho + 2 T_P^{\rho\mu} n_{\mu}\right) \nonumber \\
	&= n^{\rho} + N_{3,1} \left( 3 n^\rho + 2 n^\rho \right),
\end{align}
which entails that $N_{3,1} = -\frac{1}{5}$.

\subsubsection{Parity}

With the construction outlined above, the parity of the tensor basis will be determined by the total angular momentum
via the relation $P=(-1)^J$.
If one wants to construct a basis with the opposite parity, simply a factor of $\gamma_5$ is added to the
basis elements.

\subsubsection{Orthonormality}

It is practical to work with a basis that is orthonormalized, i.e.\ that each pair of basis vectors
$\tau_i^{\muonetoJ}$, $\tau_j^{\muonetoJ}$ obeys the orthonormality relation
\begin{equation}\label{eq:orthnormality-relation}
	\tr (\bar{\tau}_{i}^{\muonetoJ}\tau_{j}^{\muonetoJ}) = \delta_{ij},
\end{equation}
with projectors $\bar{\tau}_i^{\muonetoJ} = \pm \tau_i^{\muonetoJ}$; see Tables~\ref{tab:bases1} to \ref{tab:bases5} for the specific signs. 
Orthogonality is already achieved by construction since we only used orthogonal building blocks in our basis.
Normalization is realized via $N_{J,0}$ and $M_{J,0}$.
Finally, by adding factors of $i$ at appropriate places one is even able to make all dressing functions $f_i(p^2,P^2,p\cdot P)$ real.

Applying this construction to mesons with $J=0\dots 5$ and $P=\pm$ gives the orthonormal bases shown in 
Tables~\ref{tab:bases1} to \ref{tab:bases5}.

\bibliographystyle{utphys_mod}
\bibliography{references}

\providecommand{\href}[2]{#2}\begingroup\raggedright\begin{thebibliography}{10}

\bibitem{COMPASS:2025wkw}
{\bfseries COMPASS} Collaboration, G.~D. Alexeev {\em et~al.}, ``{Spectroscopy
  of Strange Mesons and First Observation of a Strange Crypto-Exotic State with
  $J^P=0^-$}'', \href{http://arxiv.org/abs/2504.09470}{{\ttfamily
  arXiv:2504.09470 [hep-ex]}}.

\bibitem{Quintans:2022utc}
{\bfseries AMBER} Collaboration, C.~Quintans, ``{The New AMBER Experiment at
  the CERN SPS}'', \href{http://dx.doi.org/10.1007/s00601-022-01769-7}{{\em Few
  Body Syst.} {\bfseries 63} no.~4, (2022) 72}.

\bibitem{Wallner:2022scd}
{\bfseries COMPASS, AMBER} Collaboration, S.~Wallner, ``{Strange-Meson
  Spectroscopy {\textendash} from COMPASS to AMBER}'',
  \href{http://dx.doi.org/10.1051/epjconf/202227403010}{{\em EPJ Web Conf.}
  {\bfseries 274} (2022) 03010},
  \href{http://arxiv.org/abs/2211.09499}{{\ttfamily arXiv:2211.09499
  [hep-ex]}}.

\bibitem{Brambilla:2014jmp}
N.~Brambilla {\em et~al.}, ``{QCD and Strongly Coupled Gauge Theories:
  Challenges and Perspectives}'',
  \href{http://dx.doi.org/10.1140/epjc/s10052-014-2981-5}{{\em Eur. Phys. J.}
  {\bfseries C74} no.~10, (2014) 2981},
\href{http://arxiv.org/abs/1404.3723}{{\ttfamily arXiv:1404.3723 [hep-ph]}}.

\bibitem{Joos:1962qq}
H.~Joos, ``{O}n the {R}epresentation theory of inhomogeneous {L}orentz groups
  as the foundation of quantum mechanical kinematics'', {\em Fortsch. Phys.}
  {\bfseries 10} (1962) 65--146.

\bibitem{Weinberg:1964cn}
S.~Weinberg, ``{F}eynman {R}ules for {A}ny {S}pin'',
\href{http://dx.doi.org/10.1103/PhysRev.133.B1318}{{\em Phys. Rev.} {\bfseries
  133} (1964) B1318--B1332}.

\bibitem{Zemach:1968zz}
C.~Zemach, ``{D}etermination of the {S}pins and {P}arities of {R}esonances'',
\href{http://dx.doi.org/10.1103/PhysRev.140.B109}{{\em Phys. Rev.} {\bfseries
  140} (1965) B109--B124}.

\bibitem{Krassnigg:2010mh}
A.~Krassnigg and M.~Blank, ``{A} covariant study of tensor mesons'',
  \href{http://dx.doi.org/10.1103/PhysRevD.83.096006}{{\em Phys. Rev.}
  {\bfseries D83} (2011) 096006},
\href{http://arxiv.org/abs/1011.6650}{{\ttfamily arXiv:1011.6650 [hep-ph]}}.

\bibitem{Fischer:2014xha}
C.~S. Fischer, S.~Kubrak, and R.~Williams, ``{M}ass spectra and {R}egge
  trajectories of light mesons in the {B}ethe-{S}alpeter approach'',
  \href{http://dx.doi.org/10.1140/epja/i2014-14126-6}{{\em Eur. Phys. J.}
  {\bfseries A50} (2014) 126},
\href{http://arxiv.org/abs/1406.4370}{{\ttfamily arXiv:1406.4370 [hep-ph]}}.

\bibitem{Godfrey:1985xj}
S.~Godfrey and N.~Isgur, ``{M}esons in a {R}elativized {Q}uark {M}odel with
  {C}hromodynamics'',
\href{http://dx.doi.org/10.1103/PhysRevD.32.189}{{\em Phys. Rev.} {\bfseries
  D32} (1985) 189--231}.

\bibitem{Ebert:2009ub}
D.~Ebert, R.~N. Faustov, and V.~O. Galkin, ``{Mass spectra and Regge
  trajectories of light mesons in the relativistic quark model}'',
  \href{http://dx.doi.org/10.1103/PhysRevD.79.114029}{{\em Phys. Rev. D}
  {\bfseries 79} (2009) 114029},
  \href{http://arxiv.org/abs/0903.5183}{{\ttfamily arXiv:0903.5183 [hep-ph]}}.

\bibitem{Qin:2011xq}
S.-x. Qin, L.~Chang, Y.-x. Liu, C.~D. Roberts, and D.~J. Wilson,
  ``{Investigation of rainbow-ladder truncation for excited and exotic
  mesons}'', \href{http://dx.doi.org/10.1103/PhysRevC.85.035202}{{\em Phys.
  Rev.} {\bfseries C85} (2012) 035202},
\href{http://arxiv.org/abs/1109.3459}{{\ttfamily arXiv:1109.3459 [nucl-th]}}.

\bibitem{Blank:2011ha}
M.~Blank and A.~Krassnigg, ``{Bottomonium in a Bethe-Salpeter-equation
  study}'', \href{http://dx.doi.org/10.1103/PhysRevD.84.096014}{{\em Phys.
  Rev.} {\bfseries D84} (2011) 096014},
\href{http://arxiv.org/abs/1109.6509}{{\ttfamily arXiv:1109.6509 [hep-ph]}}.

\bibitem{Williams:2015cvx}
R.~Williams, C.~S. Fischer, and W.~Heupel, ``{L}ight mesons in {QCD} and
  unquenching effects from the 3{PI} effective action'',
  \href{http://dx.doi.org/10.1103/PhysRevD.93.034026}{{\em Phys. Rev.}
  {\bfseries D93} no.~3, (2016) 034026},
\href{http://arxiv.org/abs/1512.00455}{{\ttfamily arXiv:1512.00455 [hep-ph]}}.

\bibitem{Cloet:2013jya}
I.~C. Cloet and C.~D. Roberts, ``{E}xplanation and {P}rediction of
  {O}bservables using {C}ontinuum {S}trong {QCD}'',
  \href{http://dx.doi.org/10.1016/j.ppnp.2014.02.001}{{\em Prog. Part. Nucl.
  Phys.} {\bfseries 77} (2014) 1--69},
\href{http://arxiv.org/abs/1310.2651}{{\ttfamily arXiv:1310.2651 [nucl-th]}}.

\bibitem{Eichmann:2016yit}
G.~Eichmann, H.~Sanchis-Alepuz, R.~Williams, R.~Alkofer, and C.~S. Fischer,
  ``{Baryons as relativistic three-quark bound states}'',
  \href{http://dx.doi.org/10.1016/j.ppnp.2016.07.001}{{\em Prog. Part. Nucl.
  Phys.} {\bfseries 91} (2016) 1--100},
\href{http://arxiv.org/abs/1606.09602}{{\ttfamily arXiv:1606.09602 [hep-ph]}}.

\bibitem{Eichmann:2025wgs}
G.~Eichmann, ``Hadron physics with functional methods'',
  \href{http://arxiv.org/abs/2503.10397}{{\ttfamily arXiv:2503.10397
  [hep-ph]}}.

\bibitem{Huber:2025cbd}
M.~Q. Huber, ``{A beginner's guide to functional methods in particle
  physics}'', \href{http://arxiv.org/abs/2510.18960}{{\ttfamily
  arXiv:2510.18960 [hep-ph]}}.

\bibitem{Huber:2020keu}
M.~Q. Huber, ``{Correlation functions of Landau gauge Yang-Mills theory}'',
  \href{http://dx.doi.org/10.1103/PhysRevD.101.114009}{{\em Phys. Rev. D}
  {\bfseries 101} no.~11, (2020) 114009},
  \href{http://arxiv.org/abs/2003.13703}{{\ttfamily arXiv:2003.13703
  [hep-ph]}}.

\bibitem{Huber:2018ned}
M.~Q. Huber, ``{Nonperturbative properties of Yang{\textendash}Mills
  theories}'', \href{http://dx.doi.org/10.1016/j.physrep.2020.04.004}{{\em
  Phys. Rept.} {\bfseries 879} (2020) 1--92},
  \href{http://arxiv.org/abs/1808.05227}{{\ttfamily arXiv:1808.05227
  [hep-ph]}}.

\bibitem{Huber:2020ngt}
M.~Q. Huber, C.~S. Fischer, and H.~Sanchis-Alepuz, ``{Spectrum of scalar and
  pseudoscalar glueballs from functional methods}'',
  \href{http://dx.doi.org/10.1140/epjc/s10052-020-08649-6}{{\em Eur. Phys. J.
  C} {\bfseries 80} no.~11, (2020) 1077},
  \href{http://arxiv.org/abs/2004.00415}{{\ttfamily arXiv:2004.00415
  [hep-ph]}}.

\bibitem{Huber:2021yfy}
M.~Q. Huber, C.~S. Fischer, and H.~Sanchis-Alepuz, ``{Higher spin glueballs
  from functional methods}'',
  \href{http://dx.doi.org/10.1140/epjc/s10052-021-09864-5}{{\em Eur. Phys. J.
  C} {\bfseries 81} no.~12, (2021) 1083},
  \href{http://arxiv.org/abs/2110.09180}{{\ttfamily arXiv:2110.09180
  [hep-ph]}}.

\bibitem{Huber:2025kwy}
M.~Q. Huber, C.~S. Fischer, and H.~Sanchis-Alepuz, ``{Apparent convergence in
  functional glueball calculations}'',
  \href{http://dx.doi.org/10.1140/epjc/s10052-025-14590-3}{{\em Eur. Phys. J.
  C} {\bfseries 85} no.~8, (2025) 859},
  \href{http://arxiv.org/abs/2503.03821}{{\ttfamily arXiv:2503.03821
  [hep-ph]}}.

\bibitem{Maris:1999nt}
P.~Maris and P.~C. Tandy, ``{Bethe-Salpeter study of vector meson masses and
  decay constants}'', \href{http://dx.doi.org/10.1103/PhysRevC.60.055214}{{\em
  Phys. Rev.} {\bfseries C60} (1999) 055214},
\href{http://arxiv.org/abs/nucl-th/9905056}{{\ttfamily arXiv:nucl-th/9905056
  [nucl-th]}}.

\bibitem{Qin:2011dd}
S.-x. Qin, L.~Chang, Y.-x. Liu, C.~D. Roberts, and D.~J. Wilson, ``{Interaction
  model for the gap equation}'',
  \href{http://dx.doi.org/10.1103/PhysRevC.84.042202}{{\em Phys. Rev.}
  {\bfseries C84} (2011) 042202},
\href{http://arxiv.org/abs/1108.0603}{{\ttfamily arXiv:1108.0603 [nucl-th]}}.

\bibitem{Mitter:2014wpa}
M.~Mitter, J.~M. Pawlowski, and N.~Strodthoff, ``{C}hiral symmetry breaking in
  continuum {QCD}'', \href{http://dx.doi.org/10.1103/PhysRevD.91.054035}{{\em
  Phys. Rev.} {\bfseries D91} (2015) 054035},
\href{http://arxiv.org/abs/1411.7978}{{\ttfamily arXiv:1411.7978 [hep-ph]}}.

\bibitem{Aguilar:2024ciu}
A.~C. Aguilar, M.~N. Ferreira, B.~M. Oliveira, J.~Papavassiliou, and G.~T.
  Linhares, ``{Infrared properties of the quark-gluon vertex in general
  kinematics}'', \href{http://dx.doi.org/10.1140/epjc/s10052-024-13605-9}{{\em
  Eur. Phys. J. C} {\bfseries 84} no.~11, (2024) 1231},
  \href{http://arxiv.org/abs/2408.15370}{{\ttfamily arXiv:2408.15370
  [hep-ph]}}.

\bibitem{Gao:2021wun}
F.~Gao, J.~Papavassiliou, and J.~M. Pawlowski, ``Fully coupled functional
  equations for the quark sector of qcd'',
  \href{http://dx.doi.org/10.1103/PhysRevD.103.094013}{{\em Phys. Rev. D}
  {\bfseries 103} no.~9, (2021) 094013},
  \href{http://arxiv.org/abs/2102.13053}{{\ttfamily arXiv:2102.13053
  [hep-ph]}}.

\bibitem{Chang:2011ei}
L.~Chang and C.~D. Roberts, ``Tracing masses of ground-state light-quark
  mesons'', \href{http://dx.doi.org/10.1103/PhysRevC.85.052201}{{\em Phys. Rev.
  C} {\bfseries 85} (2012) 052201},
  \href{http://arxiv.org/abs/1104.4821}{{\ttfamily arXiv:1104.4821 [nucl-th]}}.

\bibitem{Ferreira:2026gbe}
M.~N. Ferreira, A.~S. Miramontes, J.~M. Morgado, and J.~Papavassiliou, ``{Light
  mesons in the symmetric-vertex approximation}'',
  \href{http://arxiv.org/abs/2604.07221}{{\ttfamily arXiv:2604.07221
  [hep-ph]}}.

\bibitem{Bali:2000gf}
G.~S. Bali, ``{QCD forces and heavy quark bound states}'',
  \href{http://dx.doi.org/10.1016/S0370-1573(00)00079-X}{{\em Phys. Rept.}
  {\bfseries 343} (2001) 1--136},
  \href{http://arxiv.org/abs/hep-ph/0001312}{{\ttfamily arXiv:hep-ph/0001312}}.

\bibitem{Brambilla:1999xf}
N.~Brambilla, A.~Pineda, J.~Soto, and A.~Vairo, ``{Potential NRQCD: An
  Effective theory for heavy quarkonium}'',
  \href{http://dx.doi.org/10.1016/S0550-3213(99)00693-8}{{\em Nucl. Phys. B}
  {\bfseries 566} (2000) 275},
  \href{http://arxiv.org/abs/hep-ph/9907240}{{\ttfamily arXiv:hep-ph/9907240}}.

\bibitem{Brambilla:2000gk}
N.~Brambilla, A.~Pineda, J.~Soto, and A.~Vairo, ``{The QCD potential at
  O(1/m)}'', \href{http://dx.doi.org/10.1103/PhysRevD.63.014023}{{\em Phys.
  Rev. D} {\bfseries 63} (2001) 014023},
  \href{http://arxiv.org/abs/hep-ph/0002250}{{\ttfamily arXiv:hep-ph/0002250}}.

\bibitem{Koma:2006si}
Y.~Koma, M.~Koma, and H.~Wittig, ``{Nonperturbative determination of the QCD
  potential at O(1/m)}'',
  \href{http://dx.doi.org/10.1103/PhysRevLett.97.122003}{{\em Phys. Rev. Lett.}
  {\bfseries 97} (2006) 122003},
\href{http://arxiv.org/abs/hep-lat/0607009}{{\ttfamily arXiv:hep-lat/0607009
  [hep-lat]}}.

\bibitem{Koma:2006fw}
Y.~Koma and M.~Koma, ``{Spin-dependent potentials from lattice QCD}'',
  \href{http://dx.doi.org/10.1016/j.nuclphysb.2007.01.033}{{\em Nucl. Phys.}
  {\bfseries B769} (2007) 79--107},
\href{http://arxiv.org/abs/hep-lat/0609078}{{\ttfamily arXiv:hep-lat/0609078
  [hep-lat]}}.

\bibitem{Eichberg:2024svw}
M.~Eichberg and M.~Wagner, ``{Computing $1/m_Q$ and $1/m_Q^2$ corrections to
  the static potential with lattice gauge theory using gradient flow}'',
  \href{http://dx.doi.org/10.22323/1.466.0117}{{\em PoS} {\bfseries
  LATTICE2024} (2025) 117}, \href{http://arxiv.org/abs/2411.11640}{{\ttfamily
  arXiv:2411.11640 [hep-lat]}}.

\bibitem{Bali:2005fu}
{\bfseries SESAM} Collaboration, G.~S. Bali, H.~Neff, T.~Duessel, T.~Lippert,
  and K.~Schilling, ``{Observation of string breaking in QCD}'',
  \href{http://dx.doi.org/10.1103/PhysRevD.71.114513}{{\em Phys. Rev. D}
  {\bfseries 71} (2005) 114513},
  \href{http://arxiv.org/abs/hep-lat/0505012}{{\ttfamily
  arXiv:hep-lat/0505012}}.

\bibitem{Greensite:2011zz}
J.~Greensite, \href{http://dx.doi.org/10.1007/978-3-642-14382-3}{{\em {An
  introduction to the confinement problem}}}, vol.~821.
\newblock 2011.

\bibitem{Cucchieri:2017icl}
A.~Cucchieri, T.~Mendes, and W.~M. Serenone, ``{Lattice Gluon Propagator and
  One-Gluon-Exchange Potential}'',
  \href{http://dx.doi.org/10.1007/s13538-019-00665-6}{{\em Braz. J. Phys.}
  {\bfseries 49} no.~4, (2019) 548--563},
  \href{http://arxiv.org/abs/1704.08288}{{\ttfamily arXiv:1704.08288
  [hep-lat]}}.

\bibitem{Alkofer:2006gz}
R.~Alkofer, C.~S. Fischer, and F.~J. Llanes-Estrada, ``{Dynamically induced
  scalar quark confinement}'',
  \href{http://dx.doi.org/10.1142/S021773230802700X}{{\em Mod. Phys. Lett.}
  {\bfseries A23} (2008) 1105},
\href{http://arxiv.org/abs/hep-ph/0607293}{{\ttfamily arXiv:hep-ph/0607293}}.

\bibitem{Alkofer:2008tt}
R.~Alkofer, C.~S. Fischer, F.~J. Llanes-Estrada, and K.~Schwenzer, ``{The
  Quark-gluon vertex in Landau gauge QCD: Its role in dynamical chiral symmetry
  breaking and quark confinement}'',
  \href{http://dx.doi.org/10.1016/j.aop.2008.07.001}{{\em Annals Phys.}
  {\bfseries 324} (2009) 106--172},
\href{http://arxiv.org/abs/0804.3042}{{\ttfamily arXiv:0804.3042 [hep-ph]}}.

\bibitem{West:1982bt}
G.~B. West, ``{Confinement, the Wilson Loop and the Gluon Propagator}'',
  \href{http://dx.doi.org/10.1016/0370-2693(82)90394-X}{{\em Phys. Lett. B}
  {\bfseries 115} (1982) 468--472}.

\bibitem{ParticleDataGroup:2024cfk}
{\bfseries Particle Data Group} Collaboration, S.~Navas {\em et~al.}, ``{Review
  of particle physics}'',
  \href{http://dx.doi.org/10.1103/PhysRevD.110.030001}{{\em Phys. Rev. D}
  {\bfseries 110} no.~3, (2024) 030001}.

\bibitem{Jarecke:2002xd}
D.~Jarecke, P.~Maris, and P.~C. Tandy, ``{Strong decays of light vector
  mesons}'', \href{http://dx.doi.org/10.1103/PhysRevC.67.035202}{{\em Phys.
  Rev.} {\bfseries C67} (2003) 035202},
\href{http://arxiv.org/abs/nucl-th/0208019}{{\ttfamily arXiv:nucl-th/0208019
  [nucl-th]}}.

\bibitem{Williams:2018adr}
R.~Williams, ``{Vector mesons as dynamical resonances in the Bethe--Salpeter
  framework}'', \href{http://dx.doi.org/10.1016/j.physletb.2019.134943}{{\em
  Phys. Lett. B} {\bfseries 798} (2019) 134943},
  \href{http://arxiv.org/abs/1804.11161}{{\ttfamily arXiv:1804.11161
  [hep-ph]}}.

\bibitem{Santowsky2020}
N.~Santowsky, G.~Eichmann, C.~S. Fischer, P.~C. Wallbott, and R.~Williams,
  ``{$\sigma$-meson: Four-quark versus two-quark components and decay width in
  a Bethe-Salpeter approach}'',
  \href{http://dx.doi.org/10.1103/PhysRevD.102.056014}{{\em Phys. Rev. D}
  {\bfseries 102} no.~5, (2020) 056014},
  \href{http://arxiv.org/abs/2007.06495}{{\ttfamily arXiv:2007.06495
  [hep-ph]}}.

\bibitem{Miramontes:2021xgn}
A.~S. Miramontes, H.~Sanchis~Alepuz, and R.~Alkofer, ``Elucidating the effect
  of intermediate resonances in the quark interaction kernel on the timelike
  electromagnetic pion form factor'',
  \href{http://dx.doi.org/10.1103/PhysRevD.103.116006}{{\em Phys. Rev. D}
  {\bfseries 103} no.~11, (2021) 116006},
  \href{http://arxiv.org/abs/2102.12541}{{\ttfamily arXiv:2102.12541
  [hep-ph]}}.

\bibitem{Pelaez:2015qba}
J.~R. Pelaez, ``{From controversy to precision on the sigma meson: a review on
  the status of the non-ordinary $f_0(500)$ resonance}'',
  \href{http://dx.doi.org/10.1016/j.physrep.2016.09.001}{{\em Phys. Rept.}
  {\bfseries 658} (2016) 1}, \href{http://arxiv.org/abs/1510.00653}{{\ttfamily
  arXiv:1510.00653 [hep-ph]}}.

\bibitem{Heupel:2012ua}
W.~Heupel, G.~Eichmann, and C.~S. Fischer, ``{Tetraquark Bound States in a
  Bethe-Salpeter Approach}'',
  \href{http://dx.doi.org/10.1016/j.physletb.2012.11.009}{{\em Phys. Lett.}
  {\bfseries B718} (2012) 545--549},
\href{http://arxiv.org/abs/1206.5129}{{\ttfamily arXiv:1206.5129 [hep-ph]}}.

\bibitem{Eichmann:2015cra}
G.~Eichmann, C.~S. Fischer, and W.~Heupel, ``{T}he light scalar mesons as
  tetraquarks'', \href{http://dx.doi.org/10.1016/j.physletb.2015.12.036}{{\em
  Phys. Lett.} {\bfseries B753} (2016) 282--287},
\href{http://arxiv.org/abs/1508.07178}{{\ttfamily arXiv:1508.07178 [hep-ph]}}.

\bibitem{Heupel:2014ina}
W.~Heupel, T.~Goecke, and C.~S. Fischer, ``{Beyond Rainbow-Ladder in bound
  state equations}'', \href{http://dx.doi.org/10.1140/epja/i2014-14085-x}{{\em
  Eur. Phys. J.} {\bfseries A50} (2014) 85},
\href{http://arxiv.org/abs/1402.5042}{{\ttfamily arXiv:1402.5042 [hep-ph]}}.

\bibitem{Alkofer:2008et}
R.~Alkofer, C.~S. Fischer, and R.~Williams, ``{U}({A})(1) anomaly and eta-prime
  mass from an infrared singular quark-gluon vertex'',
  \href{http://dx.doi.org/10.1140/epja/i2008-10646-x}{{\em Eur. Phys. J.}
  {\bfseries A38} (2008) 53--60},
\href{http://arxiv.org/abs/0804.3478}{{\ttfamily arXiv:0804.3478 [hep-ph]}}.

\bibitem{Fischer:2014cfa}
C.~S. Fischer, S.~Kubrak, and R.~Williams, ``{Spectra of heavy mesons in the
  Bethe-Salpeter approach}'',
  \href{http://dx.doi.org/10.1140/epja/i2015-15010-7}{{\em Eur. Phys. J.}
  {\bfseries A51} (2015) 10},
\href{http://arxiv.org/abs/1409.5076}{{\ttfamily arXiv:1409.5076 [hep-ph]}}.

\bibitem{Watson:2012ht}
P.~Watson and H.~Reinhardt, ``{Bethe-Salpeter equation at leading order in
  Coulomb gauge}'', \href{http://dx.doi.org/10.1103/PhysRevD.86.125030}{{\em
  Phys. Rev.} {\bfseries D86} (2012) 125030},
\href{http://arxiv.org/abs/1211.4507}{{\ttfamily arXiv:1211.4507 [hep-ph]}}.

\bibitem{Maris:1997tm}
P.~Maris and C.~D. Roberts, ``{Pi- and K meson Bethe-Salpeter amplitudes}'',
  \href{http://dx.doi.org/10.1103/PhysRevC.56.3369}{{\em Phys. Rev.} {\bfseries
  C56} (1997) 3369--3383},
\href{http://arxiv.org/abs/nucl-th/9708029}{{\ttfamily arXiv:nucl-th/9708029
  [nucl-th]}}.

\bibitem{Maris:1997hd}
P.~Maris, C.~D. Roberts, and P.~C. Tandy, ``{P}ion mass and decay constant'',
  \href{http://dx.doi.org/10.1016/S0370-2693(97)01535-9}{{\em Phys. Lett.}
  {\bfseries B420} (1998) 267--273},
\href{http://arxiv.org/abs/nucl-th/9707003}{{\ttfamily arXiv:nucl-th/9707003
  [nucl-th]}}.

\bibitem{Maris:2003vk}
P.~Maris and C.~D. Roberts, ``{Dyson-Schwinger equations: A Tool for hadron
  physics}'', \href{http://dx.doi.org/10.1142/S0218301303001326}{{\em Int. J.
  Mod. Phys.} {\bfseries E12} (2003) 297--365},
\href{http://arxiv.org/abs/nucl-th/0301049}{{\ttfamily arXiv:nucl-th/0301049
  [nucl-th]}}.

\bibitem{Alkofer:2003jj}
R.~Alkofer, W.~Detmold, C.~S. Fischer, and P.~Maris, ``{Analytic properties of
  the Landau gauge gluon and quark propagators}'',
  \href{http://dx.doi.org/10.1103/PhysRevD.70.014014}{{\em Phys. Rev.}
  {\bfseries D70} (2004) 014014},
\href{http://arxiv.org/abs/hep-ph/0309077}{{\ttfamily arXiv:hep-ph/0309077}}.

\bibitem{Dorkin:2013rsa}
S.~M. Dorkin, L.~P. Kaptari, T.~Hilger, and B.~Kampfer, ``{Analytical
  properties of the quark propagator from a truncated Dyson-Schwinger equation
  in complex Euclidean space}'',
  \href{http://dx.doi.org/10.1103/PhysRevC.89.034005}{{\em Phys. Rev.}
  {\bfseries C89} (2014) 034005},
\href{http://arxiv.org/abs/1312.2721}{{\ttfamily arXiv:1312.2721 [hep-ph]}}.

\bibitem{Windisch:2016iud}
A.~Windisch, ``{Analytic properties of the quark propagator from an effective
  infrared interaction model}'',
  \href{http://dx.doi.org/10.1103/PhysRevC.95.045204}{{\em Phys. Rev.}
  {\bfseries C95} no.~4, (2017) 045204},
\href{http://arxiv.org/abs/1612.06002}{{\ttfamily arXiv:1612.06002 [hep-ph]}}.

\bibitem{Schlessinger:1968spm}
L.~Schlessinger, ``{Use of Analyticity in the Calculation of Nonrelativistic
  Scattering Amplitudes}'', {\em Phys. Rev.} {\bfseries 167} no.~3, (1968)
  1411.

\bibitem{Sanchis-Alepuz:2017jjd}
H.~Sanchis-Alepuz and R.~Williams, ``{Recent developments in bound-state
  calculations using the Dyson-Schwinger and Bethe-Salpeter equations}'',
\href{http://arxiv.org/abs/1710.04903}{{\ttfamily arXiv:1710.04903 [hep-ph]}}.

\bibitem{Fierz:1939zz}
M.~Fierz, ``{Force-free particles with any spin}'', {\em Helv. Phys. Acta}
  {\bfseries 12} (1939) 3--37.

\end{thebibliography}\endgroup

\end{document}